\documentclass{article}

\usepackage{amsmath}
\usepackage{amsthm}
\usepackage{amsfonts}
\usepackage{amssymb}
\usepackage{graphicx, rotating}
\usepackage{epstopdf}
\usepackage{epsfig}
\usepackage{latexsym}
\usepackage{graphicx}
\usepackage{color}
\usepackage[dvipsnames]{xcolor}
\usepackage{amsmath,amssymb}
\usepackage{cite}
\usepackage{slashed}
\usepackage{float}
\usepackage[export]{adjustbox}
\usepackage{physics}
\usepackage{textgreek}
\usepackage[dvipsnames]{xcolor}
\definecolor{rossos}{cmyk}{0,1,1,0.55}
\definecolor{myred}{rgb}{0.7, 0, 0}
\definecolor{myblue}{rgb}{0, 0, 0.7}
\definecolor{bluscuro}{rgb}{0.15, 0.2, .85}
\definecolor{bluchiaro}{cmyk}{1,.3,0.,0.1}
\usepackage[colorlinks=True, citecolor=myred, linkcolor=myblue, urlcolor=bluscuro,linktocpage]{hyperref}
\usepackage{orcidlink}
\usepackage{authblk}

\begin{document}
	
	\title{A Material Frame: Hard Recoils from Slow Force Carriers}
	\author[1]{Francesco Serra\hspace{0.05cm}\orcidlink{0000-0002-7962-3555}\thanks{\href{mailto:fserra2@jh.edu}{fserra2@jh.edu}}}
	
	\affil[1]{\small Department of Physics \& Astronomy, The Johns Hopkins University, Baltimore, MD 21218, USA}
	
	\date{\today}
	
	\maketitle
	
	\begin{abstract}\noindent
	We modify the Standard Model by making the longitudinal component of the photon physical, propagating with small speed $c_L \ll 1$. This modification breaks gauge invariance and Lorentz symmetry, defining a preferred reference frame. 
	No new fields or scales are introduced, and the photon mass is protected by a Galilean higher-form symmetry. The theory has a smooth $c_L\to0$ limit: the new modes decouple from standard matter and ordinary gauge theory predictions are recovered.
	Experimental constraints are severe. Charged particles Cherenkov-emit these slow modes, experiencing hard recoils against the preferred frame. The recoil rate decreases with $c_L$, but the momentum transfer remains of the order of the particle's momentum. Dark matter detectors, sensitive to keV-scale nuclear recoils, imply an order-of-magnitude bound $c_L \lesssim 10^{-60}$, tens of orders of magnitude stronger than speed-difference bounds. At such values, $c_L$ is better understood as a recoil-rate factor rather than a physically relevant speed. The slow modes, practically fixed in space, make the reference frame a material medium capable of absorbing momentum from charged particles.
	\end{abstract}

	\tableofcontents
\section{Introduction}
New light degrees of freedom can arise in nature in a variety of familiar ways. They may be part of a hidden sector, coupled weakly to the Standard Model, or they may parameterize a departure from the symmetries and principles that organize known physics. Lorentz symmetry, for example, can be broken explicitly or spontaneously, defining a preferred frame and leading to additional gravitational degrees of freedom, see \textit{e.g.}, \cite{Jacobson:2000xp,Arkani-Hamed:2003pdi}. 
Gauge invariance, by contrast, is usually imposed as a requirement on massless spin-one interactions \cite{Weinberg:1964ew}, and is treated as a redundancy that cannot be deformed without spoiling the theory. Violations of gauge invariance could in principle lead to negative norm states \cite{Gupta:1949rh,Bleuler:1950cy}, and result in massive force carriers \cite{Proca:1936fbw,Stueckelberg:1957zz}, loss of predictivity at high energy \cite{LlewellynSmith:1973yud,Cornwall:1974km}, and instantaneous propagation \cite{Gabadadze:2004iv,Dvali:2005nt}.

In this work we present a new way of relaxing gauge invariance while keeping the photon gapless and the theory predictive, leading to new physical excitations that could so far have gone undetected.
This possibility arises when Lorentz symmetry is spontaneously or explicitly broken. A would-be gauge degree of freedom can be made physical and assigned a very small propagation speed in a preferred frame. Although the mode remains light and couples to matter through ordinary Standard Model interactions, it decouples as its speed is taken to zero.
The suppression is kinematical, controlled by the speed of the mode rather than by a large energy scale or a small coupling constant. In this decoupling limit, Lorentz invariant gauge theory is recovered at the level of physical observables.
As we discuss, this decoupling by low speed has distinctive observational signatures.

This mechanism is not generic. Taking the speed of a gapless excitation to zero can be singular: at fixed energy the phase space can be enhanced, soft states proliferate, and perturbation theory can break down \cite{Endlich:2010hf}. In our approach, the would-be gauge nature of the slow mode ensures that its interactions are suppressed by its small speed, so that the zero-speed limit of the theory is healthy and coincides with the original gauge theory.

We discuss this construction in the case of electromagnetism. We consider a deformation of electromagnetism in which the longitudinal component of the photon, normally removed by gauge invariance, becomes physical and propagates with speed $c_L \ll 1$. 
In a preferred frame, the deformation is $\frac{c_L^2}{2}(\vec{\nabla}\!\cdot\!\vec{A})^2 \,.$ In this preferred-frame formulation, $A_0$ is not introduced as a Lagrange multiplier but rather set to zero, so the usual Gauss-law constraint is relaxed rather than rigidly imposed. 
In this way, the new longitudinal photon obeys a dispersion relation of the form:
\begin{align}
	\omega^2 = c_L^2 k^2 \,.
\end{align}
The construction introduces neither new fields nor new mass scales, as the photon mass is protected by a residual Galilean higher-form symmetry. In our approach gauge invariance is relaxed by lifting its dynamical restriction: while in gauge theory only the two transverse components of the photon carry nonzero energy, in our model the longitudinal component propagates, but carries far less energy than transverse photons at the same wavelength.
This weak deformation of gauge invariance is possible only because of the Lorentz-violating kinematics of the new slow modes.

While the slow longitudinal photon is minimally coupled to charged matter, its kinematics result in interactions suppressed by its speed $c_L$, making the limit $c_L\to0$ smooth. The Standard Model is therefore modified by a slow, gapless force carrier behaving as a weakly coupled oscillator with very small positive energy at fixed momentum.
Propagation takes place within the usual relativistic light cone, preserving the standard causal structure.

The slow longitudinal photons have distinctive testable signatures, due to their unusual kinematics for $c_L\ll1$. Being slow and gapless, they can be Cherenkov-emitted by charged particles. The rate for this process is suppressed by $c_L$, reflecting the fact that the slow modes decouple for $c_L\to 0$. However, while an emitted slow mode carries negligible energy, $\omega=c_L k$, it carries momentum of order the momentum of the charged particle. Thus, when a charged particle emits a slow mode, it undergoes a nearly elastic recoil in the preferred frame. As $c_L$ is made smaller, this process becomes rarer, but the recoil remains hard. The signal is therefore not a small perturbation to an energy level or dispersion relation; it is a rare, hard momentum transfer to the preferred frame.

This is the main observational signature of the theory. Momentum-sensitive systems become far more powerful probes than conventional cosmic-ray energy loss or precision measurements. We estimate bounds from several such systems, finding that the strongest constraints come from dark matter direct-detection experiments, designed to detect keV-scale nuclear recoils. Existing xenon recoil searches imply the order-of-magnitude bound
$$c_L \lesssim 10^{-60} \,.$$
This bound is tens of orders of magnitude stronger than those obtained from changes in dispersion relations or from the total power that cosmic rays lose by emitting slow modes, see Table~\ref{tab:bounds}.
\begin{table}[ht]
	\centering
	\renewcommand{\arraystretch}{1.6}
	\setlength{\tabcolsep}{14pt}
	\begin{tabular}{|l|c|}
		\hline
		Muon $g-2$   & $c_L\lesssim 10^{-7}$ \\
		\hline
		Cosmic-ray power loss     & $c_L\lesssim 10^{-23}$ \\
		\hline
		Recoils: Collider beams   & $c_L\lesssim 10^{-28}$ \\
		\hline
		Recoils: Pulsar spin-down rate      & $c_L\lesssim 10^{-40}$ \\
		\hline
		Recoils: Earth heat balance      & $c_L\lesssim 10^{-45}$ \\
		\hline
		Recoils: cosmic-ray nuclei      & $c_L\lesssim 10^{-48}$ \\
		\hline
		Recoils: dark matter detection (Xe)        & $c_L\lesssim 10^{-60}$ \\
		\hline
	\end{tabular}
	\caption{Experimental order-of-magnitude bounds on the slow speed $c_L$, derived in Sec.~\ref{sec:bounds}. }\label{tab:bounds}
\end{table}

At such small values, $c_L$ no longer represents a meaningful propagation speed over laboratory or cosmological scales, and is better understood as a recoil-rate parameter. The longitudinal modes are effectively fixed in space, and their observable role is to provide a channel through which charged matter can deposit momentum into the preferred frame. In this regime the theory is experimentally equivalent to the Standard Model supplemented by a rigid, nearly transparent material reference frame, with all other Lorentz-violating effects below observable sensitivity. Gauge invariance is the limit in which this material frame becomes perfectly transparent to charged matter.

Beyond its interesting phenomenological signature and decoupling mechanism, our construction also provides a dynamical realization of gauge invariance. Gauge theory is obtained as the zero-speed limit of a well-defined theory of weakly coupled slow oscillators, subject to appropriate global symmetries. For any nonzero $c_L$, the slow longitudinal photon is a physical oscillator, and it becomes a gauge mode as $c_L\to0$. 
Since the slow modes have soft energies at fixed momentum, this limit exhibits a correspondence between gauge redundancy and infrared degeneracy. In this sense, the absence of singularities for $c_L\to0$ reflects that observables in our theory are insensitive to the growing degeneracy of states that differ by soft longitudinal quanta. In fact, we will see that quantities that have a gauge-invariant limit can be written in terms of operators dressed by the slow would-be gauge modes, reflecting a mapping between gauge invariant quantities and quantities inclusive over the slow modes.

As $c_L$ is dimensionless, our model is a marginal deformation of gauge theory, with Lorentz-violating effects controlled by the same small parameter at all scales. Together with the shift symmetry protecting the photon mass, this makes the theory technically natural in the sense relevant for Lorentz-violating effective theories \cite{Myers:2003fd,Collins:2004bp}. 
Lorentz-violating effective theories have been studied extensively both in the context of modified gravity, see, \textit{e.g.}, \cite{Will:1972zz,Jacobson:2000xp,Csaki:2000dm,Arkani-Hamed:2003pdi,Gripaios:2004ms,Jacobson:2004ts,Arkani-Hamed:2004gbh,Arkani-Hamed:2005teg,Cheng:2006us,Jacobson:2007veq,Horava:2009uw,Blas:2009qj,Blas:2009yd,Blas:2010hb,Jacobson:2010mx}, and in parameterizing deviations from Lorentz invariance in the Standard Model, see, \textit{e.g.}, \cite{Colladay:1996iz,Colladay:1998fq,Kostelecky:2002hh,Kostelecky:2003fs,Kostelecky:2008ts,Altschul:2009ae}.
The model we present extends the work in these directions by showing that a preferred frame can have non-trivial Standard Model properties, and that the Standard Model can host degrees of freedom with kinematics dramatically different compared to ordinary particles.

Although gauge invariance is usually treated as an exact redundancy, rather than as a dynamical restriction that can be weakly relaxed, this construction shows that such a relaxation can be controlled. It gives rise to a new decoupling mechanism with testable experimental signatures. Dark matter detection experiments, designed to search for new forms of matter, turn out to be the sharpest probes of whether the vacuum is truly transparent to charges.

\subsection*{Summary of results}

\paragraph{Construction and symmetries.}
In Sec.~\ref{sec:dynamics} we formulate the slow-longitudinal-photon theory canonically, using electromagnetism as a minimal example. We show that adding the longitudinal-gradient term gives a stable Hamiltonian with the usual transverse photons and an additional positive-energy longitudinal oscillator. We identify the residual Galilean higher-form symmetries that protect the photon mass in the presence of charged matter, and explain how these symmetries arise as global remnants of gauge invariance. We also give an equivalent covariant description in which Lorentz symmetry is broken spontaneously by a preferred-frame field.

\paragraph{Smooth zero-speed limit.}
In Sec.~\ref{sec:decoupling} we quantize the slow mode and establish the power counting in $c_L$. The potentially dangerous factors from slow-mode phase space and propagators are compensated by speed-suppressed interactions. The speed suppression follows from the Ward identities of the theory and can be made manifest by working with charged fields dressed by the slow longitudinal mode. We show that ordinary gauge-theory rates and cross sections are recovered as $c_L\to0$, and that the theory does not develop the zero-speed strong-coupling behavior familiar from generic slow excitations \cite{Endlich:2010hf}.

\paragraph{Observable effects and constraints.}
In Sec.~\ref{sec:effects} we compute representative effects of the slow force carrier. The leading irreducible process is Cherenkov emission of a longitudinal photon by charged matter, with rate given in Eq.~\eqref{eq:rate}. We also analyze examples illustrating the general power counting: ordinary observables receive corrections suppressed by positive powers of $c_L$, while on-shell slow-mode emission can produce hard momentum transfer.
In Sec.~\ref{sec:bounds} we derive order-of-magnitude constraints on $c_L$, with particular attention to systems that are sensitive to the recoils. The strongest constraint comes from xenon recoil searches, which gives $c_L\lesssim10^{-60}$, as summarized in Table~\ref{tab:bounds}.

\paragraph{Gauss-law violation and shadow sectors.}
In Sec.~\ref{sec:Gviol} we study the classical consequences of relaxing Gauss's law. Moving charges source small apparent charge densities along their past trajectories; these charge tracks are suppressed by $c_L^2$ and are unobservable once the recoil bounds are imposed. In Sec.~\ref{sec:shadow} we describe shadow-charge sectors in the deformed theory. Configurations that are superselected in ordinary gauge theory become coherent states of the slow longitudinal field, with a slow dynamics that becomes frozen in the strict $c_L\to0$ limit.

	\section{Slow force carriers beyond gauge theory}\label{sec:dynamics}
	In this section we introduce the new model that parameterizes the simplest deviations from gauge symmetry without the addition of any scale or new fields to the theory. For simplicity, we work in the context of electromagnetism. In this approach the would-be gauge modes become weakly dynamical, as they propagate with a very low speed $c_L$. 
	As we show, the Hamiltonian remains stable. Further, in Sec.~\ref{sec:shift}, we show that a global higher-form symmetry ensures that both ordinary photons as well as the new slow modes remain gapless. In Sec.~\ref{sec:spont}, we rewrite our theory in the language of spontaneously broken Lorentz symmetry.
	We will show in Sec.~\ref{sec:decoupling} that in the limit $c_L\to0$ one smoothly recovers gauge theory, as the new slow modes decouple.
	
	We start by examining the system from the point of view of its classical Hamiltonian, using Weyl gauge, $A_0=0$. This is a convenient approach in order to formulate the canonical quantum theory, as it leads to straightforward commutation relations and reduces as much as possible the redundancy of the system while keeping the formulation local in space, see \textit{e.g.}, \cite{DelGrosso:2025ygg}. Therefore, in our approach, the quantum theory will be defined in terms of operators $\vec{A}(\vec{x},t)\,,\,\vec{E}(\vec{x},t)$, satisfying canonical equal time commutation relations: $[\vec{A}_i(\vec{x},t),\vec{E}_j(\vec{y},t)]=i\delta_{ij}\delta^3(\vec{x}-\vec{y})$. From our point of view, since we will break Lorentz symmetry down to spatial rotations, formulating the theory without introducing $A_0$ reflects the fact that the photon should be in a representation of $SO(3)$; we will comment more on the breaking of Lorentz symmetry in Sec.~\ref{sec:slowmodes} and Sec.~\ref{sec:spont}.
	In this section, for simplicity, we will just work in terms of the classical system, leaving the quantization to Sec.~\ref{sec:decoupling}. We will indicate the classical variables by the same symbols as their quantum counterparts, with commutators replaced by canonical Poisson Brackets, as we discuss in App.~\ref{app:gauge}.
	
	In the case of gauge theory, the Hamiltonian will be given by:
	\begin{align}\label{eq:H0}
		H=\int d^3x\Bigg(\frac{1}{2}\vec{E}^2+\frac{1}{2}(\vec{\nabla}\wedge\vec{A})^2\Bigg)\;.
	\end{align}
	Crucially, this Hamiltonian does not depend on the longitudinal component of $\vec{A}$, meaning that its conjugate momentum, the longitudinal component of $\vec{E}$, satisfies a local conservation law:
	\begin{align}\label{eq:Gauss}
		\frac{d}{dt}\vec{\nabla}\!\cdot\!\vec{E}=0\;.
	\end{align}
	The fact that this equation of motion is trivial tells us that the system has fewer physical degrees of freedom than canonical variables. 
	The only physical information regarding the longitudinal components of $\vec{A}$ and $\vec{E}$ is the initial condition we chose when solving Eq.~\eqref{eq:Gauss}. Classically, the longitudinal component of $\vec{A}$ can be fixed arbitrarily without changing time evolution, which makes it a gauge redundancy. This will be reflected in the quantum theory, where one defines a physical Hilbert space by discarding gauge dependent states, see \textit{e.g.} \cite{Henneaux:1992ig}. We review the canonical treatment of gauge theory in terms of the Hamiltonian in App.~\ref{app:gauge}.
	
	Usually, in this approach, one can then build a path integral over the variables $\vec{A}$ and $\vec{E}$, integrate over the conjugate momenta $\vec{E}$ and obtain a Lagrangian formulation of the theory. In this procedure, the variable $A_0$ can be reintroduced as a Lagrange multiplier that imposes Gauss's law, picking an explicit solution to Eq.~\eqref{eq:Gauss}. While the choice of setting $A_0=0$ has no physical consequences when the theory is gauge invariant, it implicitly defines a physically preferred frame as soon as we add departures from gauge theory.

	\subsection{Slow modes in the neighborhood of gauge theory}\label{sec:slowmodes}
	While our standard treatment of gauge theory is complex and rich in unphysical ingredients, the dynamical content of gauge invariance is in fact rather simple: only two polarizations of the photon carry nonzero energy. This point of view invites the possibility of slightly lifting the dynamical restriction implied by gauge invariance, as we now show.
	
	We start by asking what is a simple departure from the theory above, Eq.~\eqref{eq:H0}, in which gauge symmetry is broken. To answer this, we can demand that the theory we formulate does not introduce new fields nor energy scales, retaining its locality and symmetry under translations and rotations. Moreover, in order to have similar time-evolution to the gauge theory, we can demand that the Hamiltonian is only changed by a small amount.
	Following these points, we quickly find that we should remove the degeneracy of the Hamiltonian of Eq.~\eqref{eq:H0}, its lack of dependence on the longitudinal component of $\vec{A}$, while retaining a shift symmetry that keeps the theory gapless. Barring terms that are degenerate with the kinetic term of transverse photons, we are left with the following term:
	\begin{align}\label{eq:cl}
		(\vec{\nabla}\!\cdot\!\vec{A})^2\;,
	\end{align}
	multiplied by a small, dimensionless coefficient.
	As we will see, this small coefficient should be interpreted as the speed of the longitudinal mode. Our Hamiltonian should therefore be:
	\begin{align}\label{eq:Hem}
		H=\int d^3x\Bigg(\frac{1}{2}\vec{E}^2+\frac{1}{2}(\vec{\nabla}\!\wedge\!\vec{A})^2+\frac{c_L^2}{2}(\vec{\nabla}\!\cdot\!\vec{A})^2\Bigg)\;,\quad\text{with}\quad c_L\ll1\;.
	\end{align}
	In this setup, we have the following Hamilton's equations:
	\begin{align}\label{eq:ampere}
		\frac{d}{dt}{\vec{A}}=\vec{E}\;,\quad\frac{d}{dt}\vec{E}=\vec{\nabla}\!\wedge\!(\vec{\nabla}\!\wedge\!\vec{A})+c_L^2\vec{\nabla}(\vec{\nabla}\!\cdot\!\vec{A})\;.
	\end{align}
	If we decompose the fields in longitudinal and transverse components, e.g. $\vec{A}=\vec{A}_T+\vec{A}_L\,,\,\vec{E}=\vec{E}_T+\vec{E}_L$, we can eliminate the momenta to obtain:
	\begin{align}
		\frac{d^2}{dt^2}\vec{A}_T=\vec{\nabla}\!\wedge\!(\vec{\nabla}\!\wedge\!\vec{A}_T)\;,\quad \frac{d^2}{dt^2}\vec{A}_L=c_L^2\vec{\nabla}^2\vec{A}_L\;.
	\end{align}
	Therefore, this theory describes transverse gapless photons as well as a slower longitudinal photon. In the limit $c_L\to 0$, the longitudinal mode becomes non-dynamical and can be discarded, leading to the usual classical gauge theory. In fact, the local conservation law of Eq.~\eqref{eq:Gauss} is now changed to a continuity equation:
	\begin{align}\label{eq:GLmod}
		\frac{d}{dt}\vec{\nabla}\!\cdot\!\vec{E}=c_L^2\vec{\nabla}^2(\vec{\nabla}\!\cdot\!\vec{A})\;.
	\end{align}
	For $c_L\neq 0$, this equation cannot be solved by setting $\vec{\nabla}\!\cdot\vec{E}=0$ unless the right-hand side vanishes. This means that the Gauss operator $\vec{\nabla}\!\cdot\vec{E}$ is no longer a constraint. Rather, the longitudinal mode is now a slow oscillator with the following dispersion relation:
	\begin{align}\label{eq:disprelslow}
		\omega^2=c_L^2\vec{k}^2\;,
	\end{align}
	meaning that at a given momentum the new modes can only carry a much smaller amount of energy.
	
	In the limit $c_L \to 0$, the longitudinal mode becomes completely frozen for most practical purposes. Interestingly, we will see that this is an interesting regime that current experiments can constrain.
	In fact, while in this limit $c_L$ may become a totally negligible speed, we will find that charged particles still emit slow quanta with large momentum, at a rate that remains experimentally testable. Given the fact that in this limit the slow modes can carry only negligible energy, such momentum transfer will be, for the charged particle, the same as elastically scattering against the reference frame of the theory. 
	We develop this picture further in Sec.~\ref{sec:effects} and Sec.~\ref{sec:bounds}.
	
	While in the present work we do not consider gravitational effects, these can be derived by minimally coupling the Standard Model fields to the metric and introducing a khronon field, as discussed in \textit{e.g.} \cite{Blas:2010hb}. We discuss aspects of this approach in Sec.~\ref{sec:spont}, where we show how to rewrite our model in terms of spontaneously broken Lorentz symmetry.

	\subsection{Higher-form shift symmetries}\label{sec:shift}
	
	What symmetry protects the photon mass in our theory?
	In this section we see that Lorentz and gauge symmetry breaking in our theory leaves the transverse photon as well as the new longitudinal mode gapless, even when the field is coupled to matter. This is due to a residual global symmetry that acts as a shift on the longitudinal component of $\vec{A}$.
	
	For electromagnetism in the absence of matter, it is evident that our theory of Eq.~\eqref{eq:Hem} has a global shift symmetry, since the action only depends on $\vec{A}$ through its derivatives. In Hamiltonian terms, we have the following conserved charges:
	\begin{align}
		\vec{\Pi}=\int d^3x \vec{E}\;;\quad\frac{d}{dt}\vec{\Pi}=0\;,
	\end{align}
	which generate canonically constant shifts of $\vec{A}$. When the theory is coupled to charged matter fields, however, deriving a similar result requires more care.
	
	In fact, the coupling to matter breaks the shift symmetry above. However, we can still see that the modified Gauss's law of our theory gives us a handle to derive a global symmetry that keeps the photon as well as the new mode gapless.
	
	Assuming the charged matter fields to transform under a global $U(1)$ symmetry, we will have that the respective charge density operator, $J^0$, will appear in Gauss's law:
	\begin{align}\label{eq:cont}
		\frac{d}{dt}\Big(\vec{\nabla}\!\cdot\!\vec{E}-J^0\Big)=c_L^2\vec{\nabla}^2\vec{\nabla}\!\cdot\!\vec{A}\;.
	\end{align}
	As we have seen, this equation is obtained by taking the divergence of the Ampere's law. Therefore, its integral over all space is a trivially conserved quantity:
	\begin{align}\label{eq:globalcons}
		\frac{d}{dt}\int d^3x \Big(\vec{\nabla}\!\cdot\!\vec{E}-J^0\Big)=0\;.
	\end{align}
	This holds even in the presence of an explicit mass term for the photon, as this would appear as $m^2 \vec{\nabla}\!\cdot\!\vec{A}$ in the modified Gauss's law. In fact, this conserved charge is the sum of the conserved $U(1)$ charge of the matter fields, $Q_m=\int d^3x J^0$, and a boundary term, $Q_E=\int d^3x \vec{\nabla}\!\cdot\!\vec{E}$, which only generates shifts of the field at infinity. To extract interesting information from Gauss's law, we have to somehow remove the derivatives.
	This can be done, provided the fields fall off sufficiently rapidly at infinity.
	In fact, it is enough to integrate Eq.~\eqref{eq:cont} against $\vec{x}$ in such a way that the corresponding global charges generate constant shifts of $\vec{A}$:
	\begin{align}
		{\Pi}_1^i=\int d^3x\, x^i\Big(\vec{\nabla}\!\cdot\!\vec{E}-J^0\Big)\;,\quad\frac{d}{dt}{\Pi}_1^i=c_L^2\int d^3x\,x^i\,\vec{\nabla}^2\vec{\nabla}\!\cdot\!\vec{A}=0\;.
	\end{align}
	The ${\Pi}_1^i$ would not be conserved in the presence of a photon mass, as we would have $\frac{d}{dt}{\Pi}_1^i=m^2\int d^3x\,\vec{A}_i\neq0$. It is simple to verify that these global charges generate symmetry transformations that are the special case of a gauge transformation with parameter $\lambda=c_i \,x^i\,,$ and $c_i$ constants.
	In the same vein, if the total matter charge is zero, $Q_m=0$, then we have the following higher-order conserved global charge:
	\begin{align}
		\Pi_2^{ij}=\int d^3x\,x^{(i}x^{j)}\Big(\vec{\nabla}\!\cdot\!\vec{E}-J^0\Big)\;,\quad \frac{d}{dt}\Pi_2^{ij}=c_L^2\int d^3x\,x^{(i}x^{j)}\,\vec{\nabla}^2\vec{\nabla}\!\cdot\!\vec{A}=0\;,
	\end{align}
	where round brackets in the indices indicate symmetrization, implying six independent conservation laws.
	These conservation laws, similarly to before, are incompatible with a mass term for the photon.
	In summary, the charges ${\Pi}_1^i$ and $\Pi_2^{ij}$ generate Galileon-like infinitesimal transformations of $\vec{A}$:
	\begin{align}\label{eq:gal}
		{A}_i\to{A}_i+{c}_i+b_{ij}{x}^j\;,
	\end{align}
	as well as corresponding $\vec{x}$-dependent $U(1)$ transformations of the matter fields. Such transformations leave the action invariant up to total derivatives. Quantum mechanically, the operators corresponding to ${\Pi}_1^i$ and $\Pi_2^{ij}$ will commute with the Hamiltonian. 
	These transformations are the one-form equivalent of the scalar Galilean shifts studied in \cite{Nicolis:2008in}. While Galilean higher-form symmetries have been previously considered in \cite{Deffayet:2010zh,Tasinato:2014eka}, here they arise as a residual of broken gauge invariance and involve charged matter fields. It will be interesting to understand more precisely the connection with these previous works, as well as with works on higher-form symmetry more broadly, see \textit{e.g.} \cite{Gaiotto:2014kfa}.
	
	Beyond the higher-form symmetries we have just discussed, in Sec.~\ref{sec:irrelevant} we will see that an additional global symmetry constrains interactions between slow modes and photons.

	\subsection{Lorentz breaking as spontaneous breaking}\label{sec:spont} 
	It can be helpful to understand the system and its dynamics from the point of view of covariant fields, in a theory in which Lorentz symmetry is broken only by a clock/ether field's expectation value, which defines a preferred reference frame, see, \textit{e.g.}, \cite{Blas:2010hb,Jacobson:2007veq}. In practice, we introduce a preferred time direction through either a vector field or the gradient of a scalar field, $u_\mu$.
	The preferred reference frame will be defined by coordinates such that $u_\mu=(1,0,0,0)$. Defining $P_{\mu\nu}=\eta_{\mu\nu}-u_\mu u_\nu$ as the projector on the space-like directions (here in flat spacetime for simplicity), we can then express the vector $\vec{A}$ in terms of a generic four-vector field $A_\mu$:
	\begin{align}
		(0,\vec{A})=({P_{\mu}}^\nu A_\nu)|_\text{rest}\equiv(A^\perp_\mu)|_\text{rest}\;,
	\end{align}
	where by $|_\text{rest}$ we mean evaluating a quantity in the preferred reference frame.
	With this notation, we can write the action corresponding to Eq.~\eqref{eq:Hem} covariantly and in terms of a generic four-vector $A_\mu$. For instance, if we take $\vec{A}$ to couple linearly to a conserved $U(1)$ current $\vec{J}$, we find:
	\begin{align}\label{eq:act}
		S=\int d^4x \Bigg(-\frac{1}{4}\Big(\partial_\mu A^\perp_\nu-\partial_\nu A^\perp_\mu\Big)^2-\frac{c_L^2}{2}\Big(\partial^\mu A^\perp_\mu\Big)^2+A^\perp_\mu J^\mu\Bigg)\;.
	\end{align}
	If we take $c_L=0$, then the action is invariant under a generic transformation of the form $A^\perp_\mu\to A^\perp_\mu+\partial_\mu\lambda$, for an arbitrary function $\lambda$. Therefore, we recognize that the theory is just a rewriting of electromagnetism in Weyl gauge, and we can treat $A^\perp_\mu$ the same way we treat the vector potential in normal electromagnetism.
	
	Instead, when $c_L\neq0$, the presence of the projectors $P_{\mu\nu}$ is crucial, as it makes one of the equations of motion trivial, \textit{i.e.}, the one corresponding to $A_0$ in the preferred reference frame.
	This is an important point, as it allows the photon to break the Gauss's law constraint and gain a new dynamical component, as discussed in Sec.~\ref{sec:slowmodes}.
	
	We can see this more explicitly by deriving the equations of motion from Eq.~\eqref{eq:act}:
	\begin{align}\label{eq:cov}
		{P_\mu}^\rho\partial^\nu\left(\partial_\nu {P_\rho}^\sigma A_\sigma-\partial_\rho {P_\nu}^\sigma A_\sigma\right)+c_L^2{P_\mu}^\rho\partial_\rho(\partial^\nu {P_\nu}^\sigma A_\sigma)+{P_\mu}^\rho J_\rho=0\;.
	\end{align}
	Contracting these equations with $u^\mu$ gives a trivial equation. This means that the component of $A_\mu$ projected along $u^\mu$ is a redundancy, although the corresponding gauge transformations are not local $U(1)$ transformations, but simple local shifts of the field along $u^\mu$.
	When evaluated in the preferred reference frame, Eq.~\eqref{eq:cov} reduces to Eq.~\eqref{eq:ampere} in the presence of a conserved current.
	Taking the divergence of Eq.~\eqref{eq:cov} leads instead to a scalar equation that coincides with Eq.~\eqref{eq:GLmod} in the presence of a current, when evaluated in the preferred reference frame. 
	With these notations, the Faraday law and the magnetic Gauss law can be written as $\epsilon^{\mu\nu\rho\sigma}\partial_\nu\partial_\rho A^\perp_\sigma=0$.
	
	The covariant formulation is useful for identifying the symmetry structure of the theory, but the background $u^\mu$ is physical, and observables depend on contractions of $u^\mu$ with the momenta and with the laboratory four-velocity. A description in a frame highly boosted relative to the preferred frame of the theory will feature enhanced corrections to Maxwell's equations of order $c_L^2\gamma$, possibly with $\gamma \gg1$. Due to the severe bounds on $c_L$, see Table~\ref{tab:bounds}, ordinary observers will not experience significant enhancement of the slow-force effects.
	
	An advantage of writing a covariant action like Eq.~\eqref{eq:act} for a Lorentz breaking system is that the field $u_\mu$, when written in terms of a Stueckelberg scalar as in \cite{Blas:2010hb}, can be used to study whether coupling of the system to gravity introduces ghost instabilities or strong coupling. In fact, writing $u_\mu=\frac{\partial_\mu T}{\sqrt{(\partial T)^2}}$ and expanding the full action around $T=t+\pi(\vec{x},t)$, one can study the quadratic action for $\pi$ and read from there whether the theory has dangerous wrong-sign kinetic terms or other pathologies. Since Eq.~\eqref{eq:act} involves $A_\mu$ and $J^\mu$, it does not contribute to the quadratic action of $\pi$, and therefore does not imply any kinematic instability for the khronon field. 
	
	Before closing this section, note that the $c_L$ term of Eq.~\eqref{eq:act}, breaking gauge invariance, is only invariant under Lorentz transformations if $A^\perp_\mu$ transforms homogeneously under Lorentz boosts: $A^\perp_\mu\to \Lambda_\mu^\nu A^\perp_\nu$. This tells us that $u_\mu$ should have a non-homogeneous transformation, so as to cancel the non-homogeneous little group transformation of the massless vector field $A_\mu$. We leave a more detailed analysis of these transformation laws for future work.

	\section{Decoupling by low speed}\label{sec:decoupling}
	We now show that in the quantum theory, in the presence of matter, the limit $c_L\to 0$ is smooth and well defined, as the slow modes decouple from matter and one recovers gauge theory at the level of physical observables.
	In particular, after discussing how this decoupling works schematically in Sec.~\ref{sec:quant}, we make it manifest by defining charged fields dressed by the slow modes, in Sec.~\ref{sec:dressing}. Such fields, inclusive with respect to the slow modes, are best suited to capture the near gauge invariance of the system. Then, in Sec.~\ref{sec:weak}, we discuss the counting of powers of $c_L$ and show that transition probabilities coincide with those of gauge theory for $c_L\to0$. While the discussion of Sec.~\ref{sec:weak} may seem abstract, we provide many explicit examples of the power counting rules in Sec.~\ref{sec:effects}.
	We discuss technical aspects of canonical quantization and Ward identities in App.~\ref{app:canonical} and App.~\ref{app:ward}.
	
	This decoupling is non-trivial, and it is in fact not verified universally for slow modes in the limit of vanishing speed. For instance, the zero-speed limit is ill defined in the context of quantized fluids, see \cite{Endlich:2010hf}.
	As we will see, the crucial difference of our model with respect to these examples, is that the slow modes have interactions suppressed by their speed rather than their energy. 
	
	This feature is not obvious and is related to the nearly-gauge invariant structure of the theory. For instance, consider a case in which the vector $\vec{A}$ couples linearly to the conserved $U(1)$ current of matter, $\vec{J}$:
	\begin{align}\label{eq:Hint}
		H_\text{int}=\int d^3x\,\vec{A}\!\cdot\!\vec{J}\;.
	\end{align}
	If we decompose $\vec{A}$ in longitudinal and transverse components, $\vec{A}=\vec{A}_T+\vec{A}_L$, with $\vec{\nabla}\!\cdot\!\vec{A}_T=0$ and $\vec{A}_L=\vec{\nabla}\phi$, it would appear that the new slow modes in $\vec{A}_L$ couple to charged matter with the same strength as the usual photons.
	However, conservation of the $U(1)$ Noether current, $\vec{\nabla}\!\cdot\!\vec{J}=\frac{d}{dt}J^0$, will allow us to manipulate the interaction between the slow mode and matter schematically as:
	\begin{align}\label{eq:decouple}
		\vec{A}_L\!\cdot\!\vec{J}=\dot{\phi}J^0\;,
	\end{align}
	up to total derivatives.
	Then, the fact that in the free theory $\phi$ oscillates slowly, $\frac{d^2}{dt^2}\phi=c_L^2\vec{\nabla}^2\phi$, will imply that this interaction is suppressed by $c_L$ when $H_\text{int}$ is treated perturbatively.
	
	In the following, we will see that this manipulation can be carried out precisely in the quantum theory, with the slow modes interacting only through $\dot{\phi}$. As we will see, this result does not depend on whether or not the interaction with matter can be written as a linear coupling between $\vec{A}$ and the $U(1)$ current, as in Eq.~\eqref{eq:Hint}. Moreover, we will see that the presence of $\dot{\phi}$ in the interactions corresponds to a suppression by speed, rather than by energy of the slow modes.
	
	\subsection{Perturbation theory and speed suppression}\label{sec:quant}
	
	To start, we briefly discuss the quantization of the fields in the interaction picture, leaving to App.~\ref{app:canonical} a more detailed presentation. In particular, when expanding the fields in modes of momentum $\vec{k}$, and defining single particle momentum eigenstates, we can choose to use the same normalization conventions we use for relativistic particles, which has the benefit of leading to the Feynman rules we are used to. 
	In practice, we define the interaction picture operator for the longitudinal field $\vec{A}_L$ as:
	\begin{align}
		\vec{A}_L(\vec{x},t)=\int \frac{d^3k}{(2\pi)^3}\frac{1}{\sqrt{2c_L k}}\Big(i\hat{k} a_L^\dagger(\vec{k})e^{i(c_L kt-\vec{k}\cdot\vec{x})}-i\hat{k}a_L(\vec{k})e^{-i(c_L kt-\vec{k}\cdot\vec{x})}\Big)\;,
	\end{align}
	with $[a_L(\vec{k}),a_L^\dagger(\vec{q})]=(2\pi)^3\delta^3(\vec{k}-\vec{q})$. 
	Similarly, we adopt the customary normalization of single particle momentum eigenstates with positive norm: \begin{align}\label{eq:singlepart}
		\ket{k_L}={\sqrt{2c_Lk}}\,a_L^\dagger(\vec{k})\ket{0}\;,
	\end{align}
	leading to the same Feynman rules we use in relativistic field theory. 
	These choices make clear that dangerous powers of $\frac{1}{2c_Lk}$ will appear in phase space integrals for transition rates. 
	Physical quantities will be independent of our normalization conventions, and the dangerous factors of $c_L$ in the denominators of observable quantities will appear also if we follow a different normalization convention for field operators and states, see related discussions in \textit{e.g.} \cite{Endlich:2010hf}. 
	Such denominators might a priori spoil the $c_L\to 0$ limit, making the new modes strongly coupled at low momentum. Moreover, when slow modes in the final state have finite, nonzero energy, their momentum will be enhanced by $\frac{1}{c_L}$, leading in principle to a large contribution from the integration measure $d^3k$. Such effects are responsible for the pathologies found in \cite{Endlich:2010hf}.
	
	In order to show that the $c_L\to 0$ limit is well defined in our theory, we will show that the longitudinal mode only interacts through $\dot{\phi}$, as in Eq.~\eqref{eq:decouple}, see Sec.~\ref{sec:dressing}. We can understand right away why such rewriting of the interactions is helpful. In fact, if we write $\vec{A}_L=\vec{\nabla}\phi$, we have that $\dot{\phi}$ is manifestly suppressed by the speed $c_L$  (and not by the energy $c_L k$) with respect to $\vec{A}_L$:
	\begin{align}\label{eq:dotphi}
		\dot{\phi}(\vec{x},t)= -c_L\int \frac{d^3k}{(2\pi)^3}\frac{1}{\sqrt{2c_L k}}\Big(i {a_L^\dagger(\vec{k})}e^{i(c_L kt-\vec{k}\cdot\vec{x})}-i{a_L(\vec{k})}e^{-i(c_L kt-\vec{k}\cdot\vec{x})}\Big)\;,\quad \dot{\phi}(\vec{k},t)=-c_L\hat{k}\!\cdot\!\vec{A}_L(\vec{k},t)\;,
	\end{align}
	which makes it clear how to track factors of $c_L$ and $k$. 
	For instance, the matrix element $\langle 0 | \dot{\phi}(\vec{x},t) | k_L \rangle$ carries an overall factor of $c_L$, with no factor of $k$.
	Similarly, we can compare the propagator of $\vec{A}_L$ with the two point function of $\dot{\phi}$. For $\vec{A}_L$ we have the following Feynman propagator:
	\begin{align}\label{eq:propagsfig}
		\langle T\{\vec{A}_L^i(\vec{x}_1,t_1)\vec{A}_L^j(\vec{x}_2,t_2)\}\rangle=\int\frac{d^4k}{(2\pi)^4}\frac{i\hat{k}^i\hat{k}^j}{k_0^2-c_L^2\vec{k}^2+i\epsilon}e^{-ik_0(t_1-t_2)+i\vec{k}\cdot(\vec{x}_1-\vec{x}_2)}\;.
	\end{align}
	Instead, computing explicitly the time ordered product of two insertions of the operator $\dot{\phi}$ evaluated at generic spacetime points, we obtain\footnote{Following our relativistic intuition, we might have instead expected a numerator of the form $k_0^2/\vec{k}^2$. However, that would correspond to ignoring Schwinger terms coming from taking the time derivative of $\theta(t_1-t_2)$, which lead to an instantaneous effect $\frac{1}{\vec{k}^2}$. In fact, we have that: $\frac{k_0^2}{\vec{k}^2(k_0^2-c_L^2\vec{k}^2)}=\frac{c_L^2}{k_0^2-c_L^2\vec{k}^2}+\frac{1}{\vec{k}^2}$.
		The difference between this extra $\frac{1}{\vec{k}^2}$ and ordinary Schwinger terms, which are simple contact terms, is due to the non-standard commutation relations of $\phi$ with its conjugate momentum, which results from the canonical commutation relations between $\vec{A}_L$ and $\vec{E}_L$. Therefore, the correct result is simply obtained by computing the propagator of $\dot{\phi}$ directly, as in Eq.~\eqref{eq:propag}.}:
	\begin{align}\label{eq:propag}
		\langle T\{\dot{\phi}(\vec{x}_1,t_1)\dot{\phi}(\vec{x}_2,t_2)\}\rangle=\int\frac{d^4k}{(2\pi)^4}\frac{ic_L^2}{k_0^2-c_L^2\vec{k}^2+i\epsilon}e^{-ik_0(t_1-t_2)+i\vec{k}\cdot(\vec{x}_1-\vec{x}_2)}\;.
	\end{align}
	Unlike Eq.~\eqref{eq:propagsfig}, this propagator has a clear $c_L^2$ suppression, which makes it simple to estimate the contributions from virtual longitudinal modes in terms of $c_L$, see Sec.~\ref{sec:weak} below.
	
	In summary, Eq.~\eqref{eq:dotphi} and Eq.~\eqref{eq:propag} are the reasons why it is useful to rewrite all interactions of the slow modes in terms of $\dot{\phi}$, as in Eq.~\eqref{eq:decouple}.
	A first idea to do so might be using the Ward identities due to conservation of the global $U(1)$ Noether current. This approach requires careful evaluation of the contact terms implied by the Ward identities, which is in general not a simple task. We discuss this route in App.~\ref{app:ward}.
	Instead, here we present a simpler way to obtain the same result, which is dressing the matter fields with the slow modes, in a way that corresponds to gauge invariant fields in the $c_L\to 0$ limit, Sec.~\ref{sec:dressing}.
	
	We describe non-trivial aspects of counting powers of $c_L$ in generic diagrams in Sec.~\ref{sec:weak}, while we give explicit examples in Sec.~\ref{sec:effects}.
	
	\subsection{Dressing charges with slow modes}\label{sec:dressing}
	When we take $c_L\to0$, the spectrum of our theory becomes exactly degenerate, with the slow modes becoming gauge and carrying zero energy. From this point of view, it should not be a surprise that negative powers of $c_L$ appear at some stage in perturbation theory, analogously to IR divergences due to soft modes in gauge theory.\footnote{See \cite{Goldberger:2025mgb} for similar considerations in the context of quantized fluids.} 
	Following the arguments of the KLN theorem, see \textit{e.g.} \cite{Lee:1964is,Kinoshita:1962ur}, we expect that observables inclusive with respect to the slow modes will have a smooth $c_L\to0$ limit. Furthermore, in our case, the precise correspondence between soft modes and gauge redundancies indicates that any gauge invariant observable in gauge theory can be expressed as the zero-speed limit of a rate inclusive with respect to the slow modes. 
	These considerations suggest changing variables so as to reflect the KLN mechanism and make the emergent gauge invariance manifest already at the level of interactions. This is done by dressing the charged fields of the theory with slow modes.
	
	We can see this straightforwardly in the case of QED as well as scalar QED modified by the slow speed gradient of Eq.~\eqref{eq:cl}.
	Starting from QED, and expressing $\vec{A}=\vec{A}_T+\vec{A}_L$, with $\vec{A}_L=\vec{\nabla}\phi$ and $\vec{\nabla}\!\cdot\!\vec{A}_T=0$, we have the following Lagrangian:
	\begin{align}\begin{split}
			\mathcal{L}=\frac{1}{2}\left(\dot{\vec{A}}_T^2+\dot{\vec{A}}_L^2-(\vec{\nabla}\wedge\vec{A}_T)^2-c_L^2(\vec{\nabla}\!\cdot\!\vec{A}_L)^2\right)+\bar{\psi}\left(i\gamma^0\partial_t-i\vec{\gamma}\cdot\vec{\nabla}-e\vec{\gamma}\!\cdot\!\vec{A}_T-e\vec{\gamma}\!\cdot\!\vec{A}_L-m\right)\psi\;.
		\end{split}
	\end{align}
	We can consider the following dressing of the fermion operator:
	\begin{align}
		\Psi(x)=e^{-ie\phi(x)}\psi(x)\;.
	\end{align}
	In the $c_L\to 0$ limit, $\vec{A}_L$ should be an arbitrary space-dependent function, corresponding to the space-dependent residual gauge transformations of Weyl gauge. In this same limit, the fields $\Psi$ are the gauge invariant, physical fields for the fermions, dressed with longitudinal modes. In fact, a gauge transformation would act as $\psi\to e^{ie\alpha}\psi$ and $\phi\to\phi+\alpha$. 
	In our theory, $\phi$ has a non-trivial, but slow, time dependence, leading to a weak $\dot{\phi} J^0$ interaction.
	In practice we treat this field redefinition similar to what is discussed in, \textit{e.g.} \cite{Kamefuchi:1961sb,Coleman:1969sm,Callan:1969sn}. The perturbation theory in terms of $\Psi$ can be handled straightforwardly, as we have:
	\begin{align}\label{eq:dressed}\begin{split}
			\mathcal{L}=&\frac{1}{2}\dot{\vec{A}}^2-\frac{1}{2}(\vec{\nabla}\wedge\vec{A})^2-\frac{c_L^2}{2}(\vec{\nabla}\!\cdot\!\vec{A})^2+\bar{\Psi}\left(i\gamma^0\partial_t-i\vec{\gamma}\cdot\vec{\nabla}-e\vec{\gamma}\!\cdot\!\vec{A}_T-e\gamma^0\dot{\phi}-m\right)\Psi\;.
		\end{split}
	\end{align}	
	Notably, the dressing allowed us to trade the interaction $\vec{A}_L\!\cdot\!\vec{J}$ by the interaction $\dot{\phi} J^0$, as proposed in Eq.~\eqref{eq:decouple}. The relation $\vec{A}_L=\vec{\nabla}\phi$ implies that $\phi$ is expressed in terms of ${a_L(\vec{k},t)}/{|\vec{k}|}$ and its hermitian conjugate, meaning that the appearance of $\dot{\phi}$ corresponds to factors of $c_L$, rather than $c_L|\vec{k}|$ in interaction vertices, as discussed below Eq.~\eqref{eq:dotphi}.
	Comparing the matrix elements of this theory to matrix elements expressed in terms of $\psi$ excitations, we see that we are computing matrix elements summing over the slow modes that can be emitted by charged particles. 
	
	A similar approach can be taken in the case of scalar QED in the presence of the slow longitudinal mode. This example is interesting because we no longer have a linear coupling between the longitudinal mode and a conserved current, potentially leading to complications in the use of Ward identities.
	Nevertheless, it is simple to define a dressed scalar field invariant under spatial gauge transformations, leading to a manifestly weak coupling as in the case of the fermion.
	To see this, consider the following Lagrangian:
	\begin{align}
		\mathcal{L}=&\;\frac{1}{2}\dot{\vec{A}}^2-\frac{1}{2}(\vec{\nabla}\wedge\vec{A})^2-\frac{c_L^2}{2}(\vec{\nabla}\!\cdot\!\vec{A})^2+\dot{\chi}^*\dot{\chi}-(\vec{\nabla}+ie\vec{A})\chi^*(\vec{\nabla}-ie\vec{A})\chi\;.
	\end{align}
	We can redefine the scalar field as:
	\begin{align}
		\chi_\text{p}=e^{-ie\phi}\chi\;.
	\end{align}
	Then, the Lagrangian will be expressed in terms of the new physical field as:
	\begin{align}\label{eq:dressedchi}
		\mathcal{L}=&\;\frac{1}{2}\dot{\vec{A}}^2-\frac{1}{2}(\vec{\nabla}\wedge\vec{A})^2-\frac{c_L^2}{2}(\vec{\nabla}\!\cdot\!\vec{A})^2+(\partial_t-ie\dot{\phi}){\chi}_\text{p}^*(\partial_t+ie\dot{\phi}){\chi}_\text{p}-(\vec{\nabla}+ie\vec{A}_T)\chi_\text{p}^*(\vec{\nabla}-ie\vec{A}_T)\chi_\text{p}\;.
	\end{align}
	This means that again, the interactions with the would-be gauge invariant field $\chi_\text{p}$ are suppressed by speed rather than energy of the slow modes, which, as we will see, implies that the $c_L\to 0$ limit is smooth.

	\subsection{Power counting and absence of strong coupling}\label{sec:weak}
	
	Having shown that dressing the charged fields through slow modes allows us to express all the interactions of the slow modes in terms of $\dot{\phi}$, we can use Eq.~\eqref{eq:dotphi} and Eq.~\eqref{eq:propag} to estimate the contributions from the slow modes to various observables of the theory. We check many explicit examples in Sec.~\ref{sec:effects}. Here, we briefly summarize the scaling with $c_L$ of the various contributions to amplitudes and transition rates.
	
	\paragraph{External slow modes.} Consider a process involving $n$ external slow modes, with the initial state having finite total energy and momentum of order $p$ with respect to the preferred frame of the theory. We will distinguish two cases: (i) the slow modes carry finite momenta, of order $k_i\sim p$; (ii) the slow modes carry finite energy, of order $E_i\sim p$. Thanks to Eq.~\eqref{eq:dotphi}, we see that the amplitude $\mathcal{M}_n$ involving $n$ external slow modes will always have a suppression factor $c_L^n$ coming from the insertions of the operators $\dot{\phi}$.	
	In the first case, $k_i\sim p$, we have that the energy of the slow modes is very small: $E_i=c_Lk_i\ll p$. Then, when we compute the transition probability for the process, $\Gamma_n=\int d\Pi|\mathcal{M}_n|^2$, we find factors of $\frac{1}{E_i}\sim\frac{1}{c_Lp}$ in the phase space integral measure $d\Pi$.\footnote{For the sake of clarity, we give the explicit expression for $d\Pi$, in the case in which the $n$ slow modes are in the final state. Taking the initial state to have total four-momentum $p_\text{tot}$ and the final state to include additional $m$ speed-one modes with momenta $q_1\,,\dots\,q_m$, the phase space integration measure will have the form:
			$$
					d\Pi=\frac{1}{\Phi_\text{in}}\frac{d^3k_1}{(2\pi)^32c_Lk_1}\dots\frac{d^3k_n}{(2\pi)^32c_Lk_n}\frac{d^3q_1}{(2\pi)^32E_{q1}}\dots\frac{d^3q_m}{(2\pi)^32E_{qm}}(2\pi)^4\delta^4(p_\text{tot}-k_\text{tot}-q_\text{tot})\,,
			$$
			with $E_{q1}\,,\dots,E_{qm}$ finite, $q_\text{tot}\,,\,k_\text{tot}$ defined by summing over the corresponding four-momenta, and $\Phi_\text{in}$ the flux normalization factor for the initial state.} 
	However, since $|\mathcal{M}_n|^2\sim c_L^{2n}$, we find that the transition probability is suppressed as:
	\begin{align}\label{eq:gammasup}
		\Gamma_n=\int d\Pi|\mathcal{M}|^2\sim c_L^n\;.
	\end{align}
	The counting becomes more complex in the second case, when the slow modes carry finite energy, $E_i\sim p$. We discuss it in detail in Sec.~\ref{sec:decay}. Carrying finite energy, the slow modes will have large momentum: $k_i\sim\frac{p}{c_L}$. This leads to an enhancement in the final state phase space volume, leading to potential factors of $\int d^3k_i\sim k_i^3\sim \frac{p^3}{c_L^3}$. However, if the total momentum of the scattering is finite, of order $p$, the amplitude will have propagators that carry large momentum, leading to powers of $\frac{1}{(k_i+O(p))}\sim\frac{c_L}{p}$ in $\mathcal{M}_n$ and resulting in an additional suppression with respect to the previous case, see \textit{e.g.}, Eq.~\eqref{eq:decaymatter}. Accounting for momentum conservation, as we discuss in Sec.~\ref{sec:decay}, the transition probability is still overall suppressed by positive powers of $c_L$, similarly to Eq.~\eqref{eq:gammasup}.
	
	\paragraph{Internal slow modes.} While the propagator of Eq.~\eqref{eq:propag} has an explicit factor $c_L^2$, we should worry about contributions in which $k_0^2\ll c_L^2\vec{k}^2$. Such contributions appear to be unsuppressed: $\frac{c_L^2}{k_0^2-c_L^2\vec{k}^2}\simeq-\frac{1}{\vec{k}^2}$, comparable with the contribution from the usual photon polarization. 
	This situation can arise in two cases: as a result of integration over a loop momentum that flows through the propagator, or as a result of the kinematics of the initial state.
	In the first case, we simply see that the region of loop integral in which $k_0^2\lesssim c_L^2\vec{k}^2$ will have measure of order $c_Lk$, leading to a suppressed contribution of order $c_L$. This suppression can also be seen by taking a residue around $k_0=\pm c_Lk$, as such residue is of the order of $c_L^2/c_L=c_L$.
	
	In the second case, when the propagator is not part of a loop, we see that $k_0^2\lesssim c_L^2\vec{k}^2$ requires an extreme tuning of the kinematics of the scattering.
	In fact, our theory breaks relativity, and the propagator of Eq.~\eqref{eq:propag} is computed in the vacuum's preferred rest frame. Therefore, the order one corrections above can only appear when a scattering has negligible energy transfer with respect to the preferred frame. A scattering process which has zero energy transfer in a lab frame moving with speed $\beta$ with respect to the preferred reference frame will correspond, in our computations, to a process with energy-momentum transfer that is boosted with respect to the lab frame. This boost, by a factor $\beta$, corresponds to a propagator suppressed by $\frac{c_L^2}{\beta^2}$ whenever the lab frame is sufficiently boosted, $c_L\ll\beta$:
	\begin{align}\label{eq:boostprop}
		\frac{c_L^2}{(\beta q_\text{ext})^2-c_L^2\vec{q}_\text{ext}^{\;2}+i\epsilon}\sim\frac{c_L^2}{\beta^2}\frac{1}{\vec{q}_\text{ext}^{\;2}}\;.
	\end{align}
	In particular, if the wavefunction of the system we observe has some spread in its center-of-mass momentum, the scattering may be dominated by parts of the wave function with $\beta\gg c_L$, when $c_L$ is very small. We discuss similar considerations in Sec.~\ref{sec:coulomb}.

Concluding, we have shown that the slow modes have interactions suppressed by their speed $c_L$. Corrections from virtual slow modes are suppressed by $c_L$ in practically all observations, and so are the rates for all the processes involving external slow modes. Perturbation theory is well behaved for small $c_L$, rendering the $c_L\to0$ limit smooth. 
In the next section, we will see explicit examples of the $c_L$ scaling and study the phenomenology of the slow forces.

	\section{Effects from slow forces}\label{sec:effects}
	
	In this section we compute observable effects due to the interactions between charged particles and the slow longitudinal modes. These computations offer practical examples for the power counting derived so far, and single out the effects that can be tested experimentally, which we discuss in Sec.~\ref{sec:bounds}. 
	In particular, we see that a signal exists, which only becomes rarer, but not weaker, in the $c_L\to0$ limit: the hard recoil of a charged particle that emits a slow mode. The remaining effects discussed instead lead to weaker signals. We leave the discussion of interesting classical effects related to Gauss's law violation and shadow charges to Sec.~\ref{sec:Gviol} and Sec.~\ref{sec:shadow} respectively.
	
	\subsection{Hard recoils}\label{sec:recoil}
	Since the longitudinal modes are gapless and much slower than matter modes, we will have Cherenkov emission of on-shell longitudinal modes from on-shell charged matter.
	Imposing momentum conservation,
	\begin{align}\label{eq:pcons}
		\begin{pmatrix}
			\sqrt{p^2+m^2}\\
			\vec{p}
		\end{pmatrix}
		=\begin{pmatrix}\sqrt{p'^2+m^2}+c_L k\\
			\vec{p^{\,}}'+\vec{k}\end{pmatrix}\;,
	\end{align}
	we find the following conditions for the decay:
	\begin{align}\label{eq:ps}
		\cos\theta=c_L\frac{E_p}{p}+(1-c_L^2)\frac{k}{2p}\;,\quad\text{with}\quad 0\leq k\leq\frac{2}{1-c_L^2}(p-c_LE_p)\;,
	\end{align}
	where we have defined $\vec{p}\!\cdot\!\vec{k}=pk\cos\theta$ and we have written $E_p=\sqrt{p^2+m^2}$.
	Strikingly, the transferred momentum can be large with respect to $p$, since the charged particle experiences a nearly elastic scattering in the preferred frame of the theory. In the limit of $c_L\to0$, the momentum of the charged particle will be reflected with respect to a plane oriented at the angle $\theta$:
	\begin{align}\label{eq:recoilp}
		\vec{p}\,'\simeq\vec{p}-2p\cos\theta \,\hat{k}\;,
	\end{align}
	where we are neglecting terms of order $c_L$. The same kinematics has been observed in the case of phonons in quantized fluids, see \cite{Endlich:2010hf}.
	As we will now see, the emission of Cherenkov radiation has a $c_L$-suppressed rate. However, despite getting more and more rare as $c_L\to 0$, the recoil experienced by the charged particle remains large and observable regardless of $c_L$. This fact implies that the theory has a regime in which the recoils from hitting the rest frame of the universe are the only appreciable signature of breaking of Lorentz and gauge invariance, with all interactions being otherwise compatible with the Standard Model.
	The Cherenkov emission of a slow mode has the following matrix element:
	\begin{align}
		\mathcal{M}=c_L e\bar{u}(p')\gamma^0 u(p)\;,
	\end{align}
	where we are omitting spinor indices. Indicating with $\overline{|\mathcal{M}|^2}$ the spin-averaged matrix element square, we find the following rate for the process:
	\begin{align}\label{eq:rate}\begin{split}
			\Gamma=&\frac{1}{2E_p}\int\;\frac{d^3k d^3p'}{(2\pi)^6}\frac{1}{4E_p' c_Lk}(2\pi)^4\delta^4(p-p'-k)\overline{|\mathcal{M}|^2}\;=\frac{\alpha c_L}{ E_p}\int_{\frac{c_LE_p}{p}}^1\!\!\!\!\!d\cos\theta\frac{4(E_p^2-p^2\cos^2\theta)}{(1-c_L^2)^2}\\&=4\alpha c_L\Big(E_p-\frac{p^2}{3E_p}\Big)+O(c_L^2)\;,\end{split}
	\end{align}
	where we are assuming $p\geq c_L E_p$ so as to make the process possible.
	Contrary to our intuition that the Cherenkov rate of emission should increase with the speed difference, we see that in our theory, in the limit of $c_L\to 0$, the rate goes to zero. From our discussion in the previous sections, we can understand this as due to the speed suppression of the longitudinal mode's interactions.
	As we can see from Eq.~\eqref{eq:rate}, the scattering angle $\theta$ is sampled from a distribution $\sin^3\theta$, with $\theta$ between zero and $\pi/2$ in the limit of $c_L\to0$. While this probability distribution is peaked in $\theta=\frac{\pi}{2}$, it has an order one measure at $\theta$ small enough to make $\cos\theta$ order one.
	
	At leading order in $c_L$, we find the following power loss due to Cherenkov radiation:
	\begin{align}\label{eq:power}
		P_{\text{loss}}\simeq2\alpha c_L^2\frac{p}{E_p}(E_p^2+m^2)\;.
	\end{align}
	As we will discuss later, this process leads to bounds on $c_L$ due to the observation of cosmic rays, see Sec.~\ref{sec:bounds}.
	Still, such bounds will be much weaker than bounds from momentum-sensitive experiments, as the power loss becomes negligible for small $c_L$, while the momentum transfer remains appreciable.
	As we will see in Sec.~\ref{sec:bounds}, the momentum transfer results in the strongest experimental bounds on the speed $c_L$.
	
	While our discussion above treats the charged particle as an elementary object, we should understand what role the substructure of charged particles plays in the recoils. We discuss this in Sec.~\ref{sec:recoilbounds}, in the context of experimental bounds that can be derived from the hard recoils. Our conclusion is that the momentum transfer is limited by the lowest between the center-of-mass momentum, and the inverse characteristic size of the charge distribution, see Eq.~\eqref{eq:momtransf}.

	\paragraph{Relativistic recoils.}
	When a charged particle is highly boosted with respect to the preferred reference frame, the prediction of a hard recoil has a stark consequence in terms of the energies that the scattering process probes. 
	Of course, for an ultra-relativistic particle, the momentum transfer from the slow mode emission would probe the particle's internal structure to very high energy scales, requiring one to account for the UV degrees of freedom that compose the charged particle in order to predict the outcome of the process. This UV sensitivity is something we should expect, since in our theory a single particle state defines an invariant energy, its energy with respect to the preferred frame. If this energy is taken to be above the UV cutoff of the EFT describing the charged particle, the state should be considered outside of the regime of validity of the effective description, and we should expect time evolution to lead to a state with other UV excitations.
	
	To understand explicitly the extent of the UV sensitivity, we can consider a particle with high energy and momentum $E\,,\,p\gg m$ and perform a relativistic change of coordinates with boost parameters $\gamma=\frac{E}{m}$, $\beta=\frac{p}{E}$. Assuming $c_L\gamma\ll 1$ to be negligible, our standard relativistic intuition should apply to quantities described in this coordinate system, up to small corrections.
	In such coordinates, the momentum of the particle before the scattering is $(m,0)$. Then, the momentum transfer in these coordinates will be:
	\begin{align}\label{eq:boost}
		(p'-p)^\mu_\text{COM}=\gamma\begin{pmatrix}
			\beta k\\-\vec{k}
		\end{pmatrix}+O(c_L\gamma)
		\;=O\left({\gamma^2}{m}\right)\;.
	\end{align}
	For large $\gamma$, this momentum transfer will probe the UV structure of the bound state. While we will leave more estimates to Sec.~\ref{sec:bounds}, we can consider two relevant examples. For instance, for a proton with TeV energy, our estimate indicates that the recoil could probe the proton's structure to a scale up to $10^6\,\Lambda_\text{QCD}$, a regime of deep inelastic scattering. Even more, we can consider cosmic rays with energy of order $10^{10}\,\text{GeV}$. Then our naive analysis gives a comoving energy-momentum transfer up to order $10^{20}\,\Lambda_\text{QCD}$, in principle sensitive to physics above the Planck scale, which we cannot describe consistently.

	This seems an irreducible feature of our theory: extremely boosted particles may scatter on the rest frame and excite UV degrees of freedom which would be otherwise inaccessible.
	This UV sensitivity reflects the fact that a boosted particle provides on its own, in the presence of the slow modes, a physical energy scale, much like the invariant center-of-mass energy for scattering in a Lorentz invariant theory.
	Energy and momentum conservation will constrain the products of such scattering, with the slow mode carrying negligible energy in the preferred reference frame.
	We will discuss some of the experimental bounds implied by these considerations in Sec.~\ref{sec:bounds}.
	
	\subsection{Coulomb scattering}\label{sec:coulomb}
	The scattering between two charged particles will receive a contribution from tree-level exchange of a longitudinal mode, which will modify the potential between charges. We can compute such contribution in the preferred reference frame following Eq.~\eqref{eq:propag}:
	\begin{align}
		\mathcal{M}=\frac{i e^2 c_L^2}{\Delta E^2-c_L^2\Delta\vec{k}^2}\;,
	\end{align}
	where $\Delta E$ and $\Delta\vec{k}$ are energy and momentum exchanged between the two particles, as measured in the preferred frame. For an elastic scattering, such as electron-to-electron, in the center-of-mass frame of the system we have zero energy transfer: $\Delta E_{CM}=0$. However, if the center of mass is boosted, the exchanged energy in the preferred frame will be enhanced by a boost factor, as in Eq.~\eqref{eq:boostprop}.
	The situation is similar when we compute the contribution to the Coulomb potential between two charges. This potential is usually thought as obtained by taking the static limit in which the charges do not move, and zero energy is exchanged. However, in practice, charges do move and we have to account for relativistic corrections to the static limit, even if the system's center of mass were at rest with respect to the preferred frame. For instance, we can think of the Coulomb potential computed to study atomic spectra. This can be understood as obtained in NRQED, see \textit{e.g.}, \cite{Caswell:1985ui,Pineda:1997bj}. In this approach one treats electron and nucleus as non-relativistic, integrating out potential photons with momenta comparable with $m_e$, but neglecting photons that carry energy and momentum of order $\alpha m_e$, which contribute to higher order in the non-relativistic expansion. In the same vein, we should expand the contribution of the longitudinal mode for $\Delta E\lesssim \alpha m_e$. However, this expansion cannot be carried out consistently if $c_L$ is smaller than $\alpha$, as $c_L^2\Delta\vec{k}^2$ will be smaller than the typical $\Delta E$. Therefore, the appropriate expansion is in terms of $c_L^2/\alpha^2$, which will be small and lead to perturbative corrections to the observed atomic spectra. In practice, we have:
	\begin{align}\label{eq:deltaalpha}
		\frac{\alpha c_L^2}{\Delta E^2-c_L^2\Delta\vec{k}^2}\simeq \frac{\alpha}{\Delta\vec{k}^2}\frac{c_L^2}{\alpha^2}+O(c_L^4/\alpha^4)\;,
	\end{align}
	leading to relative corrections of the order $c_L^2/\alpha^2$ to the measured parameters of atomic physics. Since we expect most of the observed atoms to be boosted with respect to the preferred frame of the theory with a speed of the order of $\beta\sim 10^{-3}$, such boost effect leads to an energy contribution in the propagator that is small compared to the non-relativistic corrections due to electromagnetic interaction.
	
	\subsection{Matter annihilation into slow modes}\label{sec:decay}
	As a consequence of minimal coupling, charged particles can annihilate into 
	longitudinal modes. Considering for instance an initial state with two fermions, energy conservation implies that the longitudinal modes will be produced almost exactly back-to-back with momenta $\vec{k}_1\,,\,\vec{k}_2$ that are $\frac{1}{c_L}$ enhanced with respect to those of the incoming fermions $\vec{p}_1\,,\,\vec{p}_2$:
	\begin{align}\label{eq:momcons}
		|\vec{k}_1|+|\vec{k}_2|=\frac{1}{c_L}\left(\sqrt{m^2+\vec{p}_1^2}+\sqrt{m^2+\vec{p}_2^2}\right)\;,\quad\vec{k}_1+\vec{k}_2=\vec{p}_1+\vec{p}_2\;.
	\end{align}
	This means that the fermion propagator carries a large momentum of order $\frac{1}{c_L}p$, leading to an additional suppression in the amplitude. We can therefore estimate the amplitude for this process to be at most of order:
	\begin{align}
		\mathcal{M}_{2\to2}\sim \alpha c_L^3\;,
	\end{align}
	with cancellations between different diagrams potentially leading to further parametric suppression in $c_L$.
	To verify our estimates, we should examine the amplitude carefully. This is obtained from the sum of two diagrams and takes the form:
	\begin{align}\label{eq:decaymatter}
		\mathcal{M}_{2\to2}=ie^2c_L^2\bar{v}(p_2)\gamma^0\Bigg(\frac{\slashed{p}_1-\slashed{k}_1+m}{(p_1-k_1)^2-m^2}+\frac{\slashed{p}_1-\slashed{k}_2+m}{(p_1-k_2)^2-m^2}\Bigg)\gamma^0u(p_1)\;,
	\end{align}
	where $k_a=(c_L|\vec{k}_a|,\vec{k}_a)$ for $a=1,2$. When we account for the scaling in Eq.~\eqref{eq:momcons}, each diagram gives a contribution of order $\alpha c_L^3$. However, the total amplitude displays an even greater suppression, $\mathcal{M}_{2\to2}\sim \alpha c_L^4$, as the $\slashed{k}$ terms in the numerator of Eq.~\eqref{eq:decaymatter} cancel up to corrections of order $p$ when we impose momentum conservation.
	Therefore, computing the spin-averaged square amplitude, we find $\overline{|\mathcal{M}_{2\to2}|^2}\sim\alpha^2c_L^8$.
	We can similarly estimate the corresponding cross section to scale as:
	\begin{align}
		\sigma_2\sim\alpha^2 c_L^6\;,
	\end{align}
	as we expect the integral over phase space to give an enhancement of order $d\Pi\sim\frac{d^3k_1}{c_Lk_1}\frac{d^3k_2}{c_Lk_2}\delta^4(p_1+p_2-k_1-k_2)\sim\frac{p^2}{c_L^2}$. In the absence of cancellations between the diagrams, we would have simply estimated $\sigma_2\lesssim \alpha^2 c_L^4$. While integration over phase space might lead to further cancellations that are not captured by dimensional analysis, this argument shows clearly that annihilation of matter particles into longitudinal modes is extra suppressed compared to other processes. 
	For annihilation into $n$ longitudinal modes, we can give a conservative estimate by neglecting the possible cancellations, leading to cross sections at most of the order:
	\begin{align}
		\sigma_n\lesssim\alpha^n c_L^{n+2}\;.
	\end{align}
	Cancellations between diagrams, as the one discussed above, and in the phase space integrals may lead to further parametric suppression.
	Note that the inverse process in which two longitudinal modes annihilate to produce matter requires tuning the dynamics of the initial state in such a way that the longitudinal modes have almost opposite (and large) momenta.
	
	It is instructive to produce similar estimates in the case of scalar QED. While in this case interactions have up to two slow modes per vertex, each propagator carries momentum square. Only small regions of the final state phase space correspond to an amplitude that does not have propagator suppression, \textit{e.g.}, if the slow modes at the same vertices are emitted back to back. Most of the integration volume is instead penalized by the propagators, since $\frac{d^3k}{k^4}\sim c_L$. In the end, the small regions in which pairs of slow modes are emitted back to back dominate the integral, with a cross section of the order of $\sigma_n\sim \alpha^n c_L^n$, up to cancellations in the sum of diagrams or in phase space integrals.
	
	\subsection{Electron self-energy}\label{Sec:loop}
	We can compute the 1-loop correction to the electron propagator due to an off-shell longitudinal mode. We have the following:
	\begin{align}
		-i\Sigma^L(\slashed{p})=&(-ie)^2\int\frac{d^4k}{(2\pi)^4}\frac{c_L^2}{k_0^2-c_L^2\vec{k}^2+i\epsilon}\frac{\gamma^0(\slashed{p}-\slashed{k}+m)\gamma^0}{(p-k)^2-m^2+i\epsilon}\nonumber\\=&\frac{ie^2 c_L^2}{(4\pi)^2}\int_0^1dx\frac{\log(\Delta-i\epsilon)}{(1-(1-c_L^2)x)^{3/2}}\left(x\gamma^0p_0+\frac{x\,c_L^2}{1-(1-c_L^2)x}\vec{\gamma}\!\cdot\!\vec{p}+m\right)\;,
	\end{align}
	with $\Delta=(1-x)m^2-E_p^2 x(1-x)+c_L^2p^2\frac{(1-x)x}{1-(1-c_L^2)x}$. While this integral is difficult to compute, one can verify numerically that its imaginary part coincides with the Cherenkov rate computed above. Moreover, the whole integral is suppressed as $c_L$, leading only to a weak running of the speed of electrons: 
	\begin{align}
		\Delta c_e\sim -\alpha c_L \log\Big(\frac{\mu}{c_L p}\Big)\;,
	\end{align}
	where we are neglecting order one factors.
	While this scaling may have been predicted using the power counting developed so far, this is a surprising result when compared to the usual expectation that the running of speed of a particle is proportional to the difference between its speed and the speed of the particles it couples to, see \textit{e.g.}, \cite{Chadha:1982qq,Giudice:2010zb}. Interestingly, the running compares the scale $\mu$ to the typical energy, rather than the momentum, of the soft mode.
	
	\subsection{Coupling to photons}\label{sec:irrelevant}
	It is interesting to understand what operators will be generated by quantum corrections, \textit{e.g.}, by loops involving four fermion propagators. Concretely, we want to understand which operators will describe interactions when the modes have energies smaller than the electron mass, $E< m$, such that the number of electrons is conserved. In this regime, the momenta of slow modes can be very large compared to $m$, but always smaller than $m/c_L$. While a careful matching is beyond the scope of this work, we can guess what operators should appear by using naive dimensional analysis and symmetry arguments.
	
	To start, we can inspect the loop integral in a specific situation, in order to understand how the amplitude should scale in terms of $c_L$ in this regime. Consider for instance a process in which two transverse photons of momentum $p_1\,,\,p_2$, with finite energy $E<m$, are converted into two slow modes with momentum $k_1\,,\,k_2$, with large spatial components of the order of $E/c_L$. We will have a matrix element of the form:
	\begin{align}
		\mathcal{M}^{\mu\nu}_{2\gamma\to2\phi}\sim e^4c_L^2\int\frac{d^4q}{(2\pi)^4}\frac{\Tr\Big(\gamma^\mu(\slashed{q}+m)\gamma^\nu(\slashed{q}-\slashed{p}_1+m)\gamma^0(\slashed{p}_1+\slashed{p}_2-\slashed{q}+m)\gamma^0(\slashed{q}-\slashed{k}_2+m)\Big)}{(q^2-m^2)((q-p_1)^2-m^2)((q-p_1-p_2)^2-m^2)((q-k_2)^2-m^2)}\;,
	\end{align}
	where we are considering one representative contraction of the operators.
	We can estimate this integral by dividing it in three regions with non-trivial $c_L$ scaling: (i) $q$ much smaller than $E/c_L$, for which the only factors of $c_L$ come from the powers of $k_2$ in the integrand; (ii) $q$ such that $q-k_2$ is much smaller than $E/c_L$; (iii) generic $q$ of the order of $E/c_L$. The first region corresponds straightforwardly to an amplitude of the order of $\alpha^2 c_L^3$. The contribution from the second region can be estimated by shifting the integration variable as $q\to k_2+\ell$, with $\ell$ much smaller than $E/c_L$. Then we can again count powers of $c_L$ from the factors of $k_2$, obtaining a contribution of the order of $\alpha^2 c_L^5$. The third region instead corresponds to an integral that is overall independent of $c_L$, leading to an amplitude of the order of $\alpha^2c_L^2$, since $\Tr(\gamma^\mu\slashed{q}\gamma^\nu\slashed{q}\gamma^0\slashed{q}\gamma^0(\slashed{q}-\slashed{k}_2))$ is nonzero. This region dominates the integral, suggesting that the amplitude may be matched to operators in which the gradient of slow modes is weighted by the appropriate scale $\frac{m}{c_L}$.
	A similar estimate follows for the scattering of two slow modes to two slow modes, leading to an overall suppression of the amplitude by $\alpha^2c_L^4$ in this case.
	
	Given these simple estimates, we now want to understand which operators allowed by the symmetries of the full theory would reproduce such results, when electrons are integrated out.
	Besides the symmetries discussed in Sec.~\ref{sec:shift}, we know that such operators are restricted by charge conjugation, acting as $\vec{A}\to-\vec{A}$, and rotational invariance. 
	Moreover, we can see that when the matter charge is set to zero, $Q_m=\int d^3x J^0=0$, then the action is invariant under an additional symmetry in which $\dot{\phi}$ is shifted by a constant: $\dot{\phi}\to\dot{\phi}+c$. Indeed, this transformation changes the action by a term proportional to $Q_m$. This symmetry is useful, as it tells us that only operators that depend on $\vec{\nabla}\dot{\phi}$ will appear.\footnote{For instance, this symmetry forbids an operator of the form $\alpha^2\dot{\phi}^4$.} In order for such operators to be compatible with the scaling we find from our estimates, we need the gradients of the slow modes to appear weighted by the UV momentum scale $\frac{m}{c_L}$:
	\begin{align}\label{eq:self}
		E<m\,&:\quad \frac{\alpha^2}{m^4}(c_L\vec{\nabla}\dot{\phi})^2\dot{\vec{A}}\,^2_T\,,\quad\frac{\alpha^2}{m^4}(c_L\vec{\nabla}\dot{\phi})^2(\vec{\nabla}\wedge\vec{A}_T)^2\,,\quad \frac{\alpha^2}{m^4}(c_L\vec{\nabla}\dot{\phi})^4\;,
	\end{align}
	leading to an EFT in which speed suppression is still manifest.
	A precise matching computation of the processes considered here would be an important check for the above estimates, and would illustrate precisely the transition between the different regimes of the theory. We leave this for future work.

	\section{Experimental bounds}\label{sec:bounds}
	We now discuss some simple experimental bounds on the value of $c_L$. Since this parameter controls the breaking of Lorentz symmetry, our intuition may suggest that either precision measurements or the observation of cosmic rays should result in the strongest bounds. Instead, we find that the spontaneous near-elastic recoil of charged particles, due to the emission of a low speed mode, offers the best opportunity to constrain $c_L$, through momentum sensitive observations. Examples of such observations are: the control of losses in the beam of a collider, the angular momentum of pulsars, the observed heat flux through Earth's crust, the observation of heavy cosmic rays, and the recoil of atoms stored in dark matter detection experiments such as \cite{XENON:2018voc,XENON:2020rca,XENON:2023cxc}. As we will see, among the various options, dark matter experiments put the strongest constraint on our model: $c_L\lesssim 10^{-60}$, more than fifty orders of magnitude below bounds derived from precision measurements of the Standard Model's parameters.
	In the following we derive bounds as simple order-of-magnitude estimates, leaving a more precise analysis for future work.
	We start by discussing two usual probes of Lorentz symmetry breaking, precision measurements of the Standard Model parameters as well as cosmic-ray energy loss, and then switch our focus to bounds from observations sensitive to the momentum transfer in the hard recoils of Sec.~\ref{sec:recoil}. We leave the discussion of Gauss's law violation to Sec.~\ref{sec:Gviol}, where we estimate whether the ensuing effects would be measurable given the bounds we derive in this section.
	
	\subsection{Precision tests}
	As argued, observables probing electromagnetic interactions will receive relative changes that can be understood as a change in the fine structure constant: $\alpha+\delta\alpha$.
	
	For instance, in the case of the muon $g-2$, the longitudinal modes lead to a deviation of order $\alpha c_L$ in the predicted value. Given the bounds on the discrepancy between observed and predicted $(g-2)_\mu$, see, \textit{e.g.}, \cite{Aoyama:2020ynm,Muong-2:2021ojo,Muong-2:2025xyk}, we obtain:
	\begin{align}
	(g-2)_\mu\;\;\text{precision measurements:}\quad\quad	c_L\lesssim 10^{-7}\;.
	\end{align}
	In the context of atomic physics, instead, we expect the longitudinal mode to result in a deviation from $\alpha$ of the order of $\delta\alpha\sim\frac{c_L^2}{\alpha}$, as in Eq.~\eqref{eq:deltaalpha}. While it is not trivial to see whether observations in atomic physics can be combined to constrain $c_L$, taking an optimistic experimental precision of the order of $\delta\alpha\lesssim 10^{-12}$, see \textit{e.g.} \cite{parker2018measurement,Morel:2020dww}, we still find a constraint similar to the one from the muon anomalous magnetic moment: $c_L\lesssim 10^{-7}$.
	
	\subsection{Cosmic ray power loss}
	Strong bounds on $c_L$ come from the observation of cosmic rays of very high energy. The two effects that we should consider are: the power loss of charged particles due to Cherenkov radiation of slow longitudinal modes; the bounds due to the predicted discrepancy between speed of charged particles and photons. In fact, while we expect fermions' speed to be modified to order $\alpha c_L$, we expect photons to change their speed by $\alpha^2 c_L$.
	
	Cosmic rays of energy up to $10^{10}\,$GeV have been detected on Earth, and are thought to originate from extra-galactic sources \cite{PierreAuger:2017pzq,Anchordoqui:2018qom}. Such sources may be taken to be located at a distance of order $ 10$ Mpc from Earth. At the same time, if we imagine that cosmic rays receive most of their energy in the proximity of their source, we know that in order to travel such distances they cannot have energy much larger than $10^{10} \,$ GeV, as they would disintegrate against CMB photons \cite{Greisen:1966jv,Zatsepin:1966jv}. Therefore, it must be that the energy loss due to Cherenkov radiation experienced over a distance $\Delta L\sim10$ Mpc by a charged cosmic ray with energy of order $E\sim 10^{10}\,$ GeV should be much smaller than its own energy. Using units in which $c=1$, we have:
	\begin{align}
		\alpha c_L^2 E^2\frac{\Delta L}{E}\ll 1.
	\end{align}
	This implies the following bound on $c_L$:
	\begin{align}
	\text{Cosmic ray power loss:}\quad\quad	c_L\lesssim 10^{-23}\;.
	\end{align}
	Variations in the speed of fermions, lead instead to milder bounds from cosmic rays, similar to what discussed in, \textit{e.g.}, \cite{Coleman:1998ti,Coleman:1997xq,Giudice:2010zb}:
	\begin{align}
		\text{Fermion speed differences:}\quad\quad c_L\lesssim 10^{-13}\;.
	\end{align}
	Note that the above bound will be improved in the case of cosmic nuclei, as the rate will be enhanced by a factor between $Z$ and $Z^2$, $Z$ being the atomic number. We discuss this more in detail in the next section, where we will consider the effects of the hard recoils on cosmic nuclei, as well as other systems.
	
	\subsection{Bounds from hard recoils}\label{sec:recoilbounds}

	Surprisingly, the decoupling limit $c_L\to0$ allows for a signal that only becomes rare, rather than weak, as $c_L$ decreases. This is the near-elastic recoil experienced by charged particles when emitting an on-shell slow mode, which we discussed in Sec.~\ref{sec:recoil}. While the rate for this process becomes small, see Eq.~\eqref{eq:rate}, the momentum exchanged, with respect to the preferred frame of the theory, is of the order of the charged particle's momentum regardless of $c_L$, see Eq.~\eqref{eq:recoilp}.  
	Given this, situations in which we observe the momentum of a system with good accuracy become potential ways to constrain the recoil rate. As we will now see, such situations lead to the strongest bounds on $c_L$.
	
	In many situations of interest, as we will now see, the charged particles we observe would emit slow modes with wavelength short enough to resolve the particle's internal structure. Therefore, we should first of all understand the emission of slow modes from a bound state. Then, we will apply our understanding to various systems of interest: the beam in a collider, heavy cosmic rays composed of multiple nucleons, pulsars, the Earth, and finally atoms stored in dark matter direct detection experiments.
	
	In practice, since protons carry most of the momentum and energy in the systems we observe, we will neglect the recoil of electrons, which will emit slow modes with lower momentum compared to those emitted by protons.
	As stated earlier in this section, we will only derive bounds in terms of simple order-of-magnitude estimates.
	
		\paragraph{Recoils of bound states.}
	We start by discussing the emission of a slow mode by a non-relativistic charged bound state. In such a system, the momentum of the charged bound sub-components can be very different from the momentum of the center of mass. The Cherenkov emission of a slow mode by a bound state will be sensitive to both scales. 
	For instance, a bound system in the ground state of its internal configuration and at rest with respect to the preferred frame, cannot emit any slow mode. Such a transition is forbidden by energy and momentum conservation, as the system would have to gain non-vanishing center-of-mass momentum and kinetic energy.
	Instead, in situations in which the bound state has center-of-mass momentum $\vec{P}$ higher than the typical momenta of its charged sub-components, the slow mode emission probes the internal structure of the bound state and has momentum capped by the inverse size of the charge distribution, $\frac{1}{R_B}$.
	To understand this, we can consider the amplitude for the emission of a slow mode with momentum $k_\mu=(c_L|\vec{k}|,-\vec{k})$. According to, \textit{e.g.}, Eq.~\eqref{eq:dressed}, this will be given by the matrix element of the charge density operator's Fourier transform, between initial and final bound state:
	\begin{align}
		\mathcal{M}_B(k)=c_L	\bra{\vec{P}+\vec{k},B'}J^0(c_L{k},\vec{k})\ket{\vec{P},B}\;,
	\end{align}
	where $\ket{\vec{P},B}$ indicates the state with internal quantum numbers $B$ and center-of-mass momentum $\vec{P}$ in the preferred frame.
	This expression makes clear that the emission will be peaked at momenta $k$ not larger than the charge distribution's inverse size, $\frac{1}{R_B}$, as measured in the preferred frame. If $\vec{P}$ is smaller than such momenta, then the momentum exchange will be instead bound by $\vec{P}$, up to an order one factor, due to energy conservation. In summary, we will have parametrically:
	\begin{align}\label{eq:momtransf}
		k\sim \text{min}\left(P\,,\frac{1}{R_B}\right)\;.
	\end{align}
	Depending on the center-of-mass boost with respect to the preferred frame, the scattering might result in a transition to an excited state $B'$. If that is the case, the center-of-mass momentum will recoil inelastically.
	
Following the discussion about relativistic system in Sec.~\ref{sec:recoil}, we want to understand what momentum transfer to expect for the emission of a slow mode by a relativistic bound state. According to our standard relativistic intuition, a relativistic bound state with center-of-mass energy much greater than its mass, $E\gg M$, will appear shrunk by a factor $1/\gamma$ when seen from the preferred frame, with $\gamma=\frac{E}{M}\gg 1$. In our case, the internal structure of bound states is, to a small extent, determined by the non-relativistic interactions with the slow modes, leading to features that are not boost-invariant. Such corrections may become large for extreme boosts, but will remain negligible as long as $c_L\gamma\ll 1$. 
In this regime, we can think of the relativistic bound state as having a Lorentz invariant structure, meaning that it will emit slow modes with typical momentum of order $k\sim \frac{\gamma}{R_B}$, where $R_B$ is the charge distribution's typical length scale when measured at rest. As discussed in Sec.~\ref{sec:recoil}, see Eq.~\eqref{eq:boost}, the momentum transfer will probe the structure of the bound state to energies of order $\frac{\gamma^2}{R_B}$.
	
	\paragraph{Recoils in a collider.}
Collider beams are an environment with fairly controlled losses and momentum. The collider beam at LHC has a lifetime of order of ten hours, containing up to $N_p=10^{14}$ protons, see \textit{e.g.} \cite{Bruning:2004ej,Holzer:2005wnt}. While Cherenkov emission in our theory leads to small energy loss, it implies an order one momentum change for the charged particle. If a proton in the LHC beam emits a longitudinal mode, it will experience a recoil large enough to eject it from the beam's trajectory. We should demand that this loss is compatible with the measured beam lifetime at LHC. Since protons in the LHC are quite boosted, the scattering will probe the proton internal structure, meaning that only a part of the proton's momentum will be exchanged with the slow modes. At the level of the order of magnitude, we can conservatively take a momentum exchange between $10\,\text{GeV}$ and $10^2\,\text{GeV}$, see \textit{e.g.}, \cite{ParticleDataGroup:2024cfk}. Therefore, we will ask that  $\frac{10}{\alpha c_L \text{TeV}}\gtrsim 10\, \text{h}$, or:
	\begin{align}
	\text{LHC beam lifetime:}\quad\quad
	c_L\lesssim 10^{-28}\;.
	\end{align}
An even stronger bound may come from considering proton recoils happening at the collision points of LHC. Such recoil, exciting the proton's internal structure to hundreds of TeV, may result in an anomalous signal that could be detectable as an anomaly. While understanding the details of such signal and its detectability at LHC is beyond the scope of our present analysis, we can imagine the event to be detected with an efficiency $r$, between zero and one, to be determined. Then, we can impose that $c_L$ is small enough that no recoil occurring inside the LHC detector would be detected: $r\,\tau\,\Gamma\,N_p^\text{tot}<1$, where $\Gamma$ is the recoil rate, $N_p^\text{tot}$ the total number of protons circulated in the collider over the entire operation of the LHC, and $\tau$ the cumulative time each proton, on average, spent crossing the collision points. We can estimate the LHC to have gone through $10^4$ refills of the beam, resulting in a total number of protons $N_p^\text{tot}\sim 10^{18}$, each stored for an average $10$ h in the collider. Taking the collision region to have a size of order cm, see \cite{ATLAS:2017svb,Bruning:2004ej}, each proton has been effectively observed at the collision points for a time of order $\tau\sim10^{-2} s$. This results in a bound of the order $r\,c_L\lesssim 10^{-42}$, where again $0\leq r\leq 1$ is determined by the details of the recoil final state and of the data acquisition at LHC. In particular, we should only quote this as a potential bound at this stage, as it is not clear whether a recoil at a collision point would trigger detection, \textit{i.e.}, if $r\neq 0$.
	\paragraph{Pulsar spin-down.}
	As another example, we can consider pulsars. In this case, while we have less control over the individual protons in these stars, we know their spin-down rate with exquisite precision.
	For instance, millisecond pulsars are observed to have very stable frequency, with changes of the period $T$ of the order of $\frac{\dot{T}}{T}\lesssim10^{-20}\;\text{Hz}$, see, \textit{e.g.}, \cite{Lorimer:2008se}. Due to the coupling to the slow longitudinal modes, single protons in the pulsar will on average recoil from the emission of a longitudinal mode, resulting in an angular momentum loss rate proportional to the rate of Cherenkov emission and to the fraction of stellar mass composed by protons. On average, the angular momentum loss will be of the order: $\langle\dot{L}\rangle\sim-\Gamma N_p m_p R^2 \dot{\theta}$, where $N_p\sim10^{55}$ is the number of protons, $m_p\sim \text{GeV}$ the proton mass, $R\sim10\;\text{km}$ is the radius of the pulsar and $\dot{\theta}=\frac{2\pi}{T}\sim 10^3 \;\text{rad}\,\text{Hz}$ its angular speed. The total angular momentum of the pulsar is instead of the order of $L\sim M R^2\dot{\theta}$, with $M\sim 10^{57}\;\text{GeV}$ the pulsar mass. Since we have $|\frac{\langle\dot{L}\rangle}{L}|=\frac{\dot{T}}{T}$, we find the following constraint:
	\begin{align}
	\text{Pulsar spin-down:}\quad\quad
	c_L\lesssim 10^{-40}\;,
	\end{align}
	within an order of magnitude.
	
	\paragraph{Disintegration of cosmic-ray nuclei.}
	While power loss by cosmic rays is moderate, our discussion of momentum transfer makes it clear that a nucleus moving at relativistic speed with respect to the preferred frame of the theory would break apart following a hard recoil from the emission of a slow mode. Indeed, since the momentum exchanged is set by the inverse length-scale of the nuclear charge distribution, the center-of-mass momentum of the final state will not just be the reflection of the initial nucleus' momentum, but will be smaller in magnitude, allowing for some of the nucleus subcomponents to be ejected.
	Since there is convincing evidence of a flux of high energy cosmic-ray nuclei, we know that such nuclei have not emitted slow modes after having accelerated. This places a strong bound on $c_L$. For instance, if we impose that an iron nucleus traveling for $10\,\text{Mpc}$ with energy $E\sim 10^{10}\,\text{GeV}$ has low probability of emitting a slow mode, we obtain a bound of the order of:
	\begin{align}
		 \!\!\!\!\!\text{Cosmic rays nuclei disintegration:}\quad\quad c_L\lesssim 10^{-48}\;.
	\end{align}
	This is much stronger than the ordinary bound from cosmic-ray power loss, as the power loss is suppressed by an additional power of $c_L$, see Eq.~\eqref{eq:power}, as the slow modes carry little energy.
	\paragraph{Heating the Earth.} Knowing the heat flux radiated from Earth's surface allows one to constrain the recoil rate. Indeed, the momentum transfer from the recoils would increase the kinetic energy relative to the center of mass of the planet or star, effectively heating the system. This excess heating should not exceed the measured heat flow through the Earth crust. Such considerations have been explored before in \textit{e.g.} \cite{Mack:2007xj} to constrain dark matter interactions with nucleons. The measured internal heat flux is of the order of $h\sim40 \,\text{TW}\sim 10^{-1} \text{GeV}^2\,$, see \cite{Beardsmore2001}. In order to compute the energy injected by the recoils we should understand whether the atomic nuclei scatter coherently or incoherently. This can change the rate of recoil by a factor as large as $Z$, the atomic number of the heavy nuclei inside of Earth, as the rate may be enhanced if the form factor for coherent emission was sizable. For simplicity, we will take the form factor for the coherent emission to be small and accept that we might be underestimating the recoil rate by as much as two orders of magnitude. Then, we will simply ask that $N_p\Gamma \Delta E\lesssim h$, where $N_p\sim 10^{51}$ is the number of protons in the Earth, $\Gamma\sim \alpha c_L m_p$ is the recoil rate of Eq.~\eqref{eq:rate} for a non-relativistic proton, and $\Delta E$ is the kinetic energy that an atom gains on average after a recoil, measured in the reference frame comoving with Earth. Since atoms inside of the Earth move with a speed of around $\beta\sim 10^{-3}$ with respect to the cosmic frame, the emission of a slow mode from an atomic nucleus resolves its internal structure, implying a momentum transfer of the order of at least the inverse nuclear size. For a nucleus of a few fm radius, we will have a momentum transfer of the order of few times $10\,\text{MeV}$, implying a gain in kinetic energy of the atom of order $\Delta E\sim 10\,\text{keV}$.\footnote{The gain in internal energy of the nucleus is also constrained to have the same size, as the momentum transfer sets the budget of energy that can be transferred from the center-of-mass motion to excitations of the nuclear structure.}
	Putting the numbers together, we find:
	\begin{align}
		\text{Earth internal heating:}\quad\quad c_L\lesssim 10^{-45}\;,
	\end{align}
	where again, we should stress that our imprecise handling of form factors implies an uncertainty of around two orders of magnitude.
	Similar considerations about the heating of the Sun or of compact stars may lead to other interesting bounds, similarly to the case of dark matter, see \textit{e.g.} \cite{Lopes:2002gp,Bertone:2007ae}. While we leave more careful estimates for future work, we can roughly estimate bounds in terms of the maximum heating by recoil that would be consistent with observations. We expect such bounds to be weaker than for Earth: while for instance the Sun has roughly $10^6$ times the protons Earth has, its heat flux, the luminosity, is around $12$ orders of magnitude larger, see \textit{e.g.} \cite{Pr_a_2016}. This implies much less sensitivity to the heat generated by the recoils. For instance, accounting for the fact that the Sun is mostly composed of light elements (resulting in $\Delta E\sim 1 \,\text{keV}$), when we require the recoil-heat transfer to be a factor $\eta<1$ smaller than the Sun's luminosity, we obtain a bound of the form $c_L\lesssim 10^{-37}\eta$.
	
	\paragraph{Dark matter detectors.}
	As a final example, we can consider experiments for direct dark matter detection in which a tank filled with inert atoms is instrumented so as to be sensitive to the recoil of single atoms, which may follow the impact from a relic dark matter particle, see, \textit{e.g.} \cite{XENON:2018voc,XENON:2020rca,XENON:2023cxc,DarkSide-20k:2017zyg,PANDA-X:2024dlo,LZ:2022lsv,LZ:2024zvo}. Since the atoms used are neutral, we may worry that no Cherenkov radiation, and no spontaneous recoil take place. However, we can quickly understand that neutral atoms do Cherenkov radiate slow modes. As a matter of fact, the slow modes that can be emitted will have a wavelength much shorter than the atomic radius, and the nucleus exchanges much more momentum with the slow modes than the electrons.
	For a nucleus like Xe, having $A=131$ nucleons and $Z=54$ protons, we can expect the Cherenkov emission to be coherent, and the momentum exchanged to be at least of the order of the inverse size of the nucleus, $\sim 30$ MeV. If that is the case, then the nucleus will recoil with kinetic energy $\Delta E\sim \text{few keV}$, ionizing electrons that can be detected by the XENON experiment \cite{XENON:2018voc,XENON:2020rca,XENON:2023cxc}, as well as by the LZ experiment \cite{LZ:2022lsv,LZ:2024zvo} or by the PANDA-X apparatus \cite{PANDA-X:2024dlo}. The nuclear Cherenkov emission rate receives contributions from incoherent scattering, scaling as $Z$, as well as from coherent scattering, scaling as $Z^2$. At the level of an order-of-magnitude estimate, we have:
	\begin{align}
		\Gamma_\text{Xe}\sim (Z+Z^2|F(p)|^2)\alpha c_L m_p\;,
	\end{align}
	where $F(p)$ is the nuclear form factor for the process, evaluated at the scale of typical momentum transfer, $p\sim 30\;\text{MeV}$. Rather than carefully evaluating this form factor, we can content ourselves to derive bounds at the level of the order of magnitude, neglecting the possible coherent enhancement.
	We can take as a reference 1 ton of Xe and 1 year of exposure, see \cite{XENON:2018voc,XENON:2020rca,XENON:2023cxc}, and derive a bound on $c_L$ by knowing that no recoils have been detected: $N\Gamma_{\text{Xe}}\lesssim 10^{-7}\frac{1}{\text{s}}$, with $N\sim 10^{28}$ the number of Xe atoms. We therefore obtain the following order-of-magnitude bound:
	\begin{align}\label{eq:bestbound}
		\text{Xe nuclear recoil:}\quad\quad c_L\lesssim 10^{-60}\;.
	\end{align}
	As these experiments move forward, the exposure in tonne-years increases, improving the bound. In the coming years, exposures of $10\,$ton-yr will be reached, see \cite{XENON:2023cxc,LZ:2024zvo,PANDA-X:2024dlo}.
	Even more, the experiment DarkSide-20k aims at an exposure of $10^2\,\text{ton-yr}$ in Argon, with a sensitivity comparable to the experiments discussed above, see \cite{DarkSide-20k:2017zyg}, and the potential of leading to the strongest bounds on $c_L$.
	
	In conclusion, we see that current observations limit the possible values of $c_L$ to an extent that makes the slow modes practically frozen. Nonetheless, their interactions with charged matter remain testable through the induced recoils. In fact, for small $c_L$, the recoils become only more rare, while remaining a strong experimental signature.
	
	\section{Classical Gauss-law violation and charge tracks}\label{sec:Gviol}
	
	We now analyze the classical effects arising from the modification of Gauss's law, as in Eq.~\eqref{eq:GLmod}, finding that moving charges leave behind an apparent charge density, growing in time, on their past trajectory. Rather than deriving experimental bounds from this effect based on tests of Gauss's law, which would require a complex analysis, we only estimate the maximum size of the effects given the bounds in the previous section, finding that Gauss's law violation effects are too small to be measured if $c_L$ satisfies Eq.~\eqref{eq:bestbound}.
	
	We can study the classical Eq.~\eqref{eq:GLmod} by solving it perturbatively for small $c_L$, and using that $\frac{d}{dt}\vec{\nabla}\!\cdot\!\vec{A}=\vec{\nabla}\!\cdot\!\vec{E}$. We obtain, at leading order:
	\begin{align}\label{eq:GLmodpert}
		\vec{\nabla}\!\cdot\!\vec{E}-J^0\simeq c_L^2\int_0^t \!\!dt'\!\int^{t'}_0\!\!\!dt''\,\vec{\nabla}^2J^0(\vec{x},t'')\;,
	\end{align}
	where we are neglecting terms of order $c_L^4$.
	While the $c_L^2$ correction can be interpreted as a contribution that modifies the effective charge density, we know that it will not contribute to the total conserved charge when the fields vanish at infinity, following the conservation law of Eq.~\eqref{eq:globalcons}.
	For a charge smoothly distributed on a length-scale $L$ and static with respect to the absolute frame of the theory, the right-hand side of this equation grows quadratically with time, and is of order $\sim c_L^2\frac{t^2}{L^2}J^0(\vec{x})$. While this is an interesting result, we should consider what happens to charges that move with respect to the preferred frame. Taking for simplicity motion with constant speed along the coordinate $x_1$ and a Gaussian charge density of width $L$ and total charge $q$, $J^0=q\frac{\pi^{3/2}}{L^3}\exp\left(-\frac{(\vec{x}-\vec{v}t)^2}{2L^2}\right)$, we find that the right-hand side of Eq.~\eqref{eq:GLmodpert} still grows in time, although only linearly:
	\begin{align}\label{eq:effJ0}
		\vec{\nabla}\!\cdot\!\vec{E}-J^0\simeq \frac{c_L^2}{v^2}\frac{\pi^2q}{2L^3}\frac{x_1-{v}t}{L} f(\vec{x},t)\;,
	\end{align}
	with $f(\vec{x},t)\simeq\left(\text{erf}\left(\frac{vt-{x_1}}{\sqrt{2} L}\right)+\text{erf}\left(\frac{{x_1}}{\sqrt{2} L}\right)\right)e^{-\frac{x_2^2+x_3^2}{2L^2}}$, where we are neglecting terms that vanish for large $t$. For small $L$, the sum of erf$(\dots)$ terms in $f(\vec{x},t)$ approximates a product of Heaviside theta functions, $2\theta(vt-x_1)\theta(x_1)$. This result is surprising, as we get a linearly growing field at the past locations of charged particles, a charge track. Such charge track can lead to an apparent charge density with continuous value, which is in principle interesting in relation to direct detection searches for millicharged particles, see \textit{e.g.}, \cite{Essig:2011nj,Moore:2014yba,Magill:2018tbb,ArgoNeuT:2019ckq,Pospelov:2020ktu,Budker:2021quh,ArguellesDelgado:2021lek,Carney:2021irt,Iles:2024zka,Berlin:2025hjs}.
	Note that the corresponding contribution to Ampere's Law, of the form $c_L^2\int dt'\vec{\nabla}J^0$, does not grow in time for a moving charge.
	
	Since deriving experimental bounds on $c_L$ from millicharge searches would require a careful estimate of the signal due to the superposition of the multiple charge tracks passing through the experiment over the observation time, we here limit ourselves to understand whether such observable could offer a stronger probe of the slow modes than the hard recoils studied in Sec.~\ref{sec:recoil} and Sec.~\ref{sec:recoilbounds}.
	To do so, we can conservatively estimate the factor $\epsilon=\frac{c_L^2}{v^2}\frac{vt}{L}$. For instance, we can take a relativistic charge spread over the QCD scale, and an accretion time comparable with the lifetime of the universe. Then, in order for this factor to remain small, we will roughly need:
	\begin{align}
		c_L^2\lesssim 10^{-40}\;.
	\end{align}
	In other words, when $c_L$ satisfies the experimental constraints of Sec.~\ref{sec:recoilbounds}, we roughly get $\epsilon\lesssim 10^{-80}$. (Here we are approximating the exponents up to order one corrections.)
	This estimate suggests that the charge tracks will be difficult to test given the constraints on hard recoils. While it seems that considering a realistic ensemble of charge tracks flowing through an experiment should lead to a further suppression of the effect, \textit{e.g.}, due to partial cancellation between different charge tracks, we leave a careful modeling of the signal and a definitive answer for future work.
	
	Before closing, we can also give an estimate of the energy budget due to charge tracks in the observable universe. Since $\epsilon$ controls the field amplitude due to a charge track, the energy for a single charge track will scale as $E_{\text{track}}\sim\epsilon^2 \frac{vt}{L}\frac{\alpha}{L}$, with $\alpha$ being the fine structure constant and the factor $\frac{vt}{L}$ due to integrating over the entire charge track. This estimate neglects suppressions from partial cancellation of tracks due to particles with opposite charges as well as cosmic redshift, meaning that we are overestimating the energy $E_\text{track}$. With this estimate, since the total number of charged particles in the observable universe is comparable with $\frac{1}{\epsilon}$, we can expect a total energy budget strictly smaller than $\epsilon\frac{t}{L}\frac{\alpha}{L}$, which is insignificant for cosmology.
	
	\section{Shadow charges, relaxed}\label{sec:shadow}
	In gauge theory, the local conservation law of the Gauss's law operator, $\frac{d}{dt}\vec{\nabla}\!\cdot\!\vec{E}=0$, allows for solutions with nontrivial initial conditions for the longitudinal electric field, $\vec{\nabla}\!\cdot\!\vec{E}=\rho(\vec{x})$, in which the field responds to an arbitrary static charge density $\rho(\vec{x})$ that has no matter counterpart, a shadow charge. While shadow charges do not break Lorentz symmetry at a fundamental level, see discussions in \cite{DelGrosso:2025ygg}, they can be thought of as a classical analog of the properties of a preferred reference frame that we are proposing here. As a matter of fact, the slow modes that we have introduced can be thought of as the excitations that interpolate between different shadow charge sectors. In this section we explore this connection, studying how shadow charges are described in the presence of slow force carriers.
	
	Once the longitudinal components of the field have become slowly evolving, the theory will predict solutions that are similar to the shadow charges of gauge theory, in which however the shadow charge freely propagates with the low speed $c_L$. This can be seen from Eq.~\eqref{eq:ampere}, assigning the appropriate initial conditions $\vec{\nabla}\!\cdot\!{\vec{E}}(t=t_0)=\rho(\vec{x})$.
	Shadow charges in gauge theory are superselected quantities, and correspond to physical coherent states of the longitudinal electric field, see \textit{e.g.}~\cite{Kaplan:2023fbl,DelGrosso:2025ygg}.
	Such states can be straightforwardly adopted in our theory, with the difference that our theory will give the shadow charge a non-trivial, but slow, time evolution, $\rho(\vec{x},t)$. Such evolution will slowly spread the given initial shadow charge on the wavefront defined by the speed $c_L$.
	As an example, a state with shadow charge $\rho(\vec{x},t)$ in the interaction picture will be given by:
	\begin{align}\label{eq:shadowstate}
		\ket{\rho(t)}=\exp\Bigg(i\int d^3x\vec{A}(\vec{x},t)\!\cdot\!\vec{\nabla}\xi(\vec{x},t)\Bigg)\ket{0}\;,\quad\text{with}\quad\vec{\nabla}^2\xi=\rho(\vec{x},t)\;,
	\end{align}
	and $\ket{0}$ indicating the interaction picture vacuum state. Note that the exponential in the definition of these states can be rewritten as a Fourier transform, as $\ket{\rho(t)}=\exp\Big(-i\int \frac{d^3k}{(2\pi)^3}\phi(-\vec{k},t)\rho(\vec{k},t)\Big)\ket{0}$.
	Clearly, interactions between charged particles and shadow charges benefit from a coherent enhancement: the factor $\frac{1}{\sqrt{c_L}}$ contained in the $\vec{A}$ field in the exponential, combines with the time evolution operator to cancel the $c_L$ suppression of interactions of the longitudinal field with fermions. Due to this, charged particles react to the shadow charge as if it was a slowly moving classical charge. To see this explicitly, we can consider the non-relativistic limit of the equation of motion for a fermion in the presence of a shadow charge. Using the approximation $\psi\simeq e^{-imt}\begin{pmatrix}\chi\\\eta\end{pmatrix}$, with derivatives of $\chi$ and $\eta$ taken to be small compared to $m$, we obtain the following:
	\begin{align}
		-i\partial_t\chi(\vec{y},t)\simeq\left(\frac{\vec{\nabla}^2_y}{2m}+e\langle{\dot{\phi}(\vec{y},t)}\rangle\right)\chi(\vec{y},t)\;,\quad\eta\simeq-i\frac{\vec{\sigma}\!\cdot\!\vec{\nabla}\chi}{2m}\;,
	\end{align}
	where $\vec{\sigma}$ indicates the Pauli matrices.
	Therefore, a non relativistic, charged particle of mass $m$ follows classical trajectories derived from the following non-relativistic local Hamiltonian:
	\begin{align}
		H_{NR}=\frac{\vec{\nabla}^2_y}{2m}+e\langle{\dot{\phi}(\vec{y},t)}\rangle\;,\quad\text{with}\quad\langle{\dot{\phi}(\vec{y},t)}\rangle=\int \frac{d^3k}{(2\pi)^3}\frac{\rho_k(t)}{|\vec{k}|^2}e^{i\vec{k}\cdot\vec{y}}\;.
	\end{align}
	This means that the charged particle feels a Coulomb potential due to the shadow charge, as this propagates on the $c_L$ spacetime cone. For instance, in the case of $\rho(\vec{y},t_0)=Q \delta^3(\vec{y}-\vec{y}_0)$ at a given time $t_0$, we have $\langle\dot{\phi}(\vec{y},t_0)\rangle=\frac{Q}{4\pi|\vec{y}-\vec{y}_0|}$.
	
	Due to momentum conservation and momentum exchange with charged particles, the longitudinal field describing the shadow charge will also evolve to a configuration with fixed, nonzero momentum. For instance, in the case of a charged particle scattering against the shadow charge, with a momentum transfer $\Delta\vec{P}$, the shadow charge configuration will evolve to a state with expectation value  $\langle\phi\rangle\to\langle\phi\rangle'=\langle\phi\rangle+\delta\phi$, such that the total momentum of the system is constant:
	\begin{align}\label{eq:momentumtransf}
		\Delta\vec{P}_i=\int d^3x\, (\langle \vec{\nabla}_j\dot{\phi} \rangle'\langle\vec{\nabla}_i\vec{\nabla}_j\phi\rangle'-\langle \vec{\nabla}_j\dot{\phi} \rangle\langle\vec{\nabla}_i\vec{\nabla}_j\phi\rangle)=2\int d^3x\, \rho\vec{\nabla}_i\delta\phi\;.
	\end{align}
	From this, since we have $\rho=\langle\vec{\nabla}^2\dot{\phi}\rangle$ and $\rho'=\langle\vec{\nabla}^2\dot{\phi}\rangle'=\rho+\delta\rho$, we see that a finite momentum transfer $\Delta\vec{P}$ will correspond to a small change in shadow charge. This is because we will have ${\delta\rho}=O(c_L)\times\delta\phi$, since the two variations are related by a time derivative. Then, since $\Delta\vec{P}$ is linear in $\delta\phi$, it must scale as $\delta\rho/c_L$, meaning that a finite $\Delta\vec{P}$ corresponds to a $\delta\rho$ which is suppressed by a factor of $c_L$ with respect to $\rho$: 
	\begin{align}
		{\delta\rho}=O(c_L\,\rho)\;.
	\end{align}
	Specializing our analysis to the limit of $c_L=0$, we find an interesting fact: in gauge theory, any momentum transfer is entirely accounted for by the gauge component of the gauge field. In fact, for $c_L=0$, the field takes the form $\langle\vec{\nabla}^2 \phi\rangle=t\rho(\vec{x})+\lambda(\vec{x})$; writing $\rho=\vec{\nabla}^2\xi\,,\,\lambda=\vec{\nabla}^2\zeta$, the classical momentum of the field is given by:
	\begin{align}
		\vec{P}_i=\int d^3x\,\rho(t\vec{\nabla}_i\xi(\vec{x})+\vec{\nabla}_i\zeta(\vec{x}))=\int d^3x\,\rho(\vec{x})\vec{\nabla}_i\zeta(\vec{x})\;,
	\end{align}
	where we have used that first term in the integrand is a total derivative, $\int\rho\vec{\nabla}\xi=-\frac{1}{2}\int\vec{\nabla}(\vec{\nabla}\xi)^2$. In this limit, the momentum transferred to the field changes only $\lambda$, the pure gauge component of the field, while $\rho$ is superselected and unaltered by momentum transfer: $\delta\rho=0$.
	
	Given the shadow charge states, it is interesting to understand the phenomenological consequence of the interactions between longitudinal modes and charged particles as well as transverse photons. 
	In this regard, we can see that all of these interactions will conserve the total amount of shadow charge. This point is a consequence of our modified Gauss's law and of the conservation laws discussed in Sec.~\ref{sec:shift}. In fact, as we have seen in Eq.~\eqref{eq:globalcons}, the integral of $\vec{\nabla}\!\cdot\!\vec{E}-J^0$ over space is conserved. However, the integral of $J^0$ is a conserved charge on its own, meaning that the total amount of shadow charge is conserved:
	\begin{align}
		\frac{d}{dt}\int d^3x\,\rho(\vec{x},t)=0\;.
	\end{align}
	Compatibly with this condition, shadow charge distributions with large gradients, with momenta close to $m/c_L$, may give rise to non-negligible self-interactions, that we estimated in Eq.~\eqref{eq:self}. As such interactions are only relevant for short wavelength modes, they should only be important at a microscopic level.
	
	Before closing, note that superpositions of shadow charges are trivially described by superpositions of states of the form of Eq.~\eqref{eq:shadowstate}. A simple description of such superpositions is potentially useful in gravity, where it might help address the problem of time, see, \textit{e.g.}, \cite{Isham:1992ms,Kaplan:2023wyw,DelGrosso:2024gnj}.
	
	\section{Discussion and Outlook}\label{sec:conclusions}
	
	The main lesson of this work is that gauge invariance can be relaxed in a controlled way without necessarily producing a massive photon or an inconsistent theory. In the Lorentz-violating deformation studied here, the would-be longitudinal gauge mode becomes a physical, gapless excitation with speed $c_L\ll1$, while a Galilean higher-form symmetry protects the photon from acquiring a mass. The same small speed provides a decoupling mechanism: as $c_L\to0$, the longitudinal mode decouples from conserved matter currents and ordinary electrodynamics is recovered at the level of physical observables. 
	
	At finite $c_L$, however, the mode has a striking phenomenology. Charged particles can emit it with a rate suppressed by $c_L$, but when emission occurs the momentum transfer is hard. 	
	This is a qualitatively new signature of Lorentz violation: the signal becomes rare, but not weak. Recoil-sensitive observations, such as the lifetime of a collider beam, pulsar spin-down, Earth's internal heat, the observation of cosmic-ray nuclei, and dark matter detectors, become far more powerful probes than precision tests or bounds from cosmic-ray power loss.
	The resulting hierarchy of bounds spans over fifty orders of magnitude: from $c_L \lesssim 10^{-7}$ (precision tests) to $c_L \lesssim 10^{-60}$ (dark matter direct detection). 
	At such values, $c_L$ is no longer usefully interpreted as an observable propagation speed; phenomenologically, it acts as a recoil-rate parameter. The slow force carriers are practically fixed in space, and describe momentum stored in the preferred frame. In the regime consistent with observations, the theory is phenomenologically equivalent to the Standard Model supplemented by a material reference frame that can exchange momentum with charged particles. Dark matter detection experiments, designed to search for new particles, turn out to be the sharpest probes of whether the vacuum is truly transparent.
	
	Several directions remain open:
	
	\paragraph{Formal developments.} While the $c_L\to 0$ limit gives observables that coincide with those of gauge theory, it will be interesting to further explore the structural aspects of this limit, and its technical relation to the ordinary treatment of relativistic gauge theory. Similarly, it will be interesting to further investigate the spontaneous breaking of Lorentz symmetry in our model. As mentioned, our framework gives a description of gauge invariance in terms of a soft-mode degeneracy. The dressed states of Sec.~\ref{sec:dressing} are closely related to the Faddeev-Kulish dressing that resolves IR divergences in QED~\cite{Kulish:1970ut}. The connection between slow modes and soft divergences, and the role of the KLN theorem, deserve further exploration. Furthermore, the Galilean higher-form symmetries protecting the photon mass may connect to recent work on generalized global symmetries~\cite{Gaiotto:2014kfa}.
	In addition to these aspects, it will be interesting to understand more systematically the renormalization of the theory.
	
	\paragraph{Extensions to Yang-Mills and gravity.} The extension to non-abelian gauge theory presents both technical and conceptual challenges: the slow modes would carry color charge, and their self-interactions require a careful treatment. This is the object of ongoing work. Gravity presents additional puzzles. While it is clear that gravity locally deforms the rest frame, see \textit{e.g.} \cite{Kaplan:2023wyw,DelGrosso:2024gnj}, fully breaking diffeomorphism invariance may lead to ghost modes. If this is the case, any viable construction would have to control the associated instabilities or show that they are harmless, for instance if their speed is small enough. If parametrically slow positive- and negative-energy modes can be consistently described, their cosmological production could have interesting phenomenological consequences. These phenomenological aspects, as well as the general construction of a healthy theory of slow modes for gravity, remain for future study.
	
	\paragraph{Cosmological implications.} As the slow modes are emitted over the lifetime of the universe, they will form a fluid with a momentum density that can be transmitted back to charged particles. In the present work, we have not considered such effects. However, this will be an important aspect to study, as the slow modes may leave imprints on the CMB or large-scale structure. While the coupling of the slow modes to the early time Standard Model plasma needs to be understood precisely, the momentum of these slow modes will redshift to a scale comparable to the temperature of the CMB photons. 
	
	\paragraph{Experimental outlook.} Current bounds already make $c_L$ completely negligible as a speed. Yet the corresponding recoil rate remains testable, and $c_L$ continues to be relevant as a recoil rate parameter. Future dark matter experiments with lower thresholds and larger exposures will continue to probe gauge invariance in this new regime. While we have not analyzed the role of recoils on orbital electrons, that will be an observable to explore. The framework demonstrates an unexpected synergy: experiments designed to detect dark matter become precision tests of the fundamental symmetries of nature.
	Beyond the recoils, it will be interesting to study more accurately the classical Gauss's law violations discussed in Sec.~\ref{sec:Gviol} and the corresponding signals in millicharge detection experiments.
	
	\vspace{1em}
	In summary, gauge invariance can be deformed and viewed dynamically as the limit in which the vacuum is perfectly transparent to charged matter. The deformation studied here gives this transparency a finite, testable opacity: rare, hard recoils of charged particles. Whether the vacuum has such material properties is therefore a question that recoil experiments can probe.

	\section{Acknowledgments}
	I thank Loris Del Grosso, David E. Kaplan, Alessandro Podo, Surjeet Rajendran, and Riccardo Rattazzi for useful discussions and comments.
		This work was supported by the U.S. Department of Energy (DOE), Office of Science, National
		Quantum Information Science Research Centers, Superconducting Quantum Materials and Systems Center (SQMS) under Contract No. DE-AC02-07CH11359, and by the Simons Investigator Grant No. 144924. 
	
	\appendix
	
	\section{Dynamical view of gauge theory} \label{app:gauge}
	In this appendix we review the treatment of gauge invariant systems from the point of view of their canonical Hamiltonian. In particular, we discuss classical electromagnetism and its features in terms of the classical phase space (to be mapped to its Hilbert space in the quantum theory), and its canonical Poisson brackets. More details along these lines, including why it is consistent, and in fact preferable, to not use Dirac brackets, can be found in \cite{DelGrosso:2025ygg}.
	This brief review outlines the dynamical character of gauge invariance and motivates viewing gauge invariance in terms of local conservation laws, lack of dynamical freedom, and field components that carry zero energy.
	
	In its minimal formulation, electromagnetism can be written canonically in terms of a vector field $\vec{A}$ and its conjugate momentum $\vec{E}$ satisfying the commutation relations $\{\vec{A}(\vec{x}),\vec{E}(\vec{y})\}=\delta^3(\vec{x}-\vec{y})$. {Here the Poisson brackets are defined as: $\{A,B\}=\sum_n \frac{\partial A}{\partial q_n}\frac{\partial B}{\partial p_n}-\frac{\partial A}{\partial p_n}\frac{\partial B}{\partial q_n}$, where $q_n$ and $p_n$ are the canonical variables and their conjugate momenta. The subscript $n$ includes the spatial position (or the wave-number) at which the variables are evaluated.} The Hamiltonian can be taken of the following form:
	\begin{align}
		H=\int d^3x\Bigg(\frac{1}{2}\vec{E}^2+\frac{1}{2}(\vec{\nabla}\!\wedge\!\vec{A})^2\Bigg)\;.
	\end{align}
	Notably, this Hamiltonian has no dependence on the longitudinal component of $\vec{A}$, which implies that its conjugate momentum is locally conserved:
	\begin{align}\label{eq:GL}
		\frac{d}{dt}\vec{\nabla}\!\cdot\!\vec{E}=0\;.
	\end{align}
	Note that this is a combination of Hamilton's equations, obtained by taking the divergence of Ampere's law:
	\begin{align}
		\frac{d}{dt}\vec{E}=\{\vec{E},H\}=\vec{\nabla}\wedge(\vec{\nabla}\wedge\vec{A})\;.
	\end{align}
	Eq.~\eqref{eq:GL} is simple enough that it can be integrated straightforwardly, leading to:
	\begin{align}\label{eq:constr}
		\vec{\nabla}\cdot\vec{E}=\rho_s(\vec{x})\;,
	\end{align}
	with $\rho_s(\vec{x})$ a set of initial conditions, labeling the section of phase space that we choose to study. The usual Gauss's law is obtained with trivial initial conditions $\rho_s(\vec{x})=0$. Non-trivial initial conditions describe field configurations in which the electric field responds to a Coulomb field without a source, or shadow charge density, $\rho_s(\vec{x})$, see \textit{e.g.} \cite{Kaplan:2023fbl,Kaplan:2023wyw,DelGrosso:2024gnj,DelGrosso:2025ygg}. Regardless of the initial conditions $\rho_s(\vec{x})$, the solution given by Eq.~\eqref{eq:constr} is known as a constraint, since it prescribes a section of the classical phase space (and in the quantum theory, Hilbert space) that the system is constrained to live in at all times.
	Aside from telling us that the system has a simpler dynamics than we might have expected, the local conservation law of Eq.~\eqref{eq:GL} tells us that the Hamiltonian is independent of the canonical transformations generated by the locally conserved quantity $\vec{\nabla}\!\cdot\!\vec{E}$. In fact, if we define such transformations as:
	\begin{align}
		\delta_\lambda F=\lambda(\vec{x})\{F,\vec{\nabla}\!\cdot\!\vec{E}(\vec{x})\}\;,
	\end{align}
	we see that:
	\begin{align}
		\delta_\lambda H=\lambda\{H,\vec{\nabla}\!\cdot\!\vec{E}\}=-\frac{d}{dt}\vec{\nabla}\!\cdot\!\vec{E}=0\;.
	\end{align}
	This means that if we change the fields by one such transformation, with arbitrary $\lambda(\vec{x})$, time evolution is unchanged. In other words, transformations of this kind are an exact symmetry of the theory, reflecting the presence of redundant variables. It is easy to see that the redundant variables are the longitudinal modes of the vector field $\vec{A}$, as it holds:
	\begin{align}
		\delta_\lambda\vec{A}(\vec{x})=-\vec{\nabla}\lambda(\vec{x})\;.
	\end{align}
	In general, it was proven in \cite{DelGrosso:2025ygg} that local conservation laws as in Eq.~\eqref{eq:GL} correspond to gauge transformations, regardless of whether the constraints are first or second class.
	
	With this understanding, we can move towards a more familiar formulation of gauge theory. We can express the electric field in terms of $\dot{\vec{A}}$ and compute the action. Furthermore, we can introduce a Lagrange multiplier selecting our choice of initial conditions for Eq.~\eqref{eq:GL}. Calling this Lagrange multiplier $A_0$, we obtain the following action:
	\begin{align}
		S=\int d^4x\Bigg(-\frac{1}{4}F^{\mu\nu}F_{\mu\nu}+A_0\rho_s(\vec{x})\Bigg)\;,
	\end{align}
	which coincides, in the cases of $\rho_s=0$, with our usual formulation of electromagnetism. The field $A_0$ can of course be seen as a canonical variable, with conjugate momentum $\Pi_0$ subject to a new constraint, which encodes the fact that $A_0$ is not a dynamical quantity and can be changed, together with $\vec{A}$ by a gauge transformation. While $A_0$ is not strictly needed to formulate the theory, it allows to make the action manifestly Lorentz invariant, up to the dependence on the initial condition $\rho_s$. 
	
	\section{Canonical quantization}\label{app:canonical}
	In this appendix, we canonically quantize the classical model of electromagnetism in the presence of a slow longitudinal mode. We construct step by step the quantized fields in the interaction picture, outlining the various normalization choices and discussing alternatives. 
	
	Importantly, the observables of the theory by construction do not depend on the normalization convention that we pick for operators and states. Therefore, we can choose a familiar normalization convention or define a new one, with no difference in our results.
	
	We start by decomposing the canonical variables $\vec{A}\,,\,\vec{E}$ in modes {in the Schrodinger representation}. We have:
	\begin{align}
		\vec{A}(\vec{x})=\int\frac{d^3 k}{(2\pi)^3}\sum_{r=1}^3\vec{e}_r(\vec{k})\alpha_r(\vec{k})e^{i\vec{k}\cdot\vec{x}}\;,\quad \vec{E}(\vec{x})=\int\frac{d^3 k}{(2\pi)^3}\sum_{r=1}^3\vec{e}_r(\vec{k})\pi_r(\vec{k})e^{i\vec{k}\cdot\vec{x}}\;,
	\end{align}
	where we have chosen the vectors $\vec{e}_r(\vec{k})$ as a suitable basis of two transverse and a longitudinal unit vector, satisfying:
	\begin{align}
		\sum_{r=1}^3\vec{e}_{r,i}(\vec{k})\vec{e}_{r,j}(-\vec{k})=\delta_{ij}\;,\quad\sum_{i=1}^3\vec{e}_{r,i}(\vec{k})\vec{e}_{s,i}(-\vec{k})=\delta_{rs}\;.
	\end{align}
	We also have the following hermiticity conditions:
	\begin{align}
		\alpha_r(-\vec{k})=\alpha^\dagger(\vec{k})\;,\quad\pi_r(-\vec{k})=\pi_r^\dagger(\vec{k})\;,\quad \vec{e}_r(-\vec{k})=\vec{e}_r^{\;*}(\vec{k})\;.
	\end{align}
	An example of polarization vectors, for $\vec{k}=k\hat{z}$, is the following:
	\begin{align}
		\vec{e}_{1}=\frac{1}{\sqrt{2}}\Big(1,i,0\Big)\;,\quad\vec{e}_{2}=\frac{1}{\sqrt{2}}\Big(1,-i,0\Big)\;,\quad\vec{e}_{3}=\Big(0,0,i\Big)\;.
	\end{align}
	The canonical commutation relations between our operators in position space are:
	\begin{align}
		[\vec{A}_i(\vec{x}),\vec{E}_j(\vec{y})]=i\delta_{ij}\delta^3(\vec{x}-\vec{y})\;,
	\end{align}
	which imply the following rules for the mode operators:
	\begin{align}
		[\alpha_r(\vec{k}),\pi_s(\vec{q})]=i(2\pi)^3\delta_{rs}\delta^3(\vec{k}+\vec{q})\;.
	\end{align}
	Plugging this mode expansion in the Hamiltonian we obtain:
	\begin{align}
		H=\frac{1}{2}\int \frac{d^3k}{(2\pi)^3}\Bigg(\pi_{T}(\vec{k})\pi_{T}^\dagger(\vec{k})+k^2\alpha_{T}(\vec{k})\alpha_{T}^\dagger(\vec{k})+\pi_L(\vec{k})\pi_L^\dagger(\vec{k})+c_L^2k^2\alpha_L(\vec{k})\alpha_L^\dagger(\vec{k})\Bigg)\;,
	\end{align}
	where we have changed the labeling of polarization indices and we are implicitly summing over two transverse polarizations that we indicate with subscript ${}_T$.
	We can now define creation and annihilation operators, and follow the conventional normalization in our choices. For the transverse polarizations we have:
	\begin{align}
		\alpha_T(\vec{k})&=\frac{1}{\sqrt{2k}}\left(a_T(\vec{k})+a_T^\dagger(-\vec{k})\right)\;,\quad\pi_T(\vec{k})=-i\sqrt{\frac{k}{2}}\left(a_T(\vec{k})-a_T^\dagger(-\vec{k})\right)\;,\\
		a_T(\vec{k})&=\frac{1}{\sqrt{2}}\left(\sqrt{k}\,\alpha_T(\vec{k})+\frac{i}{\sqrt{k}}\pi_T(\vec{k})\right)\;,\quad a_T^\dagger(\vec{k})=\frac{1}{\sqrt{2}}\left(\sqrt{k}\,\alpha_T^\dagger(\vec{k})-\frac{i}{\sqrt{k}}\pi_T^\dagger(\vec{k})\right)\;.
	\end{align}
	We can follow the conventional normalization also in defining the longitudinal creation and annihilation operators:
	\begin{align}\label{eq:norm}
		\alpha_L(\vec{k})&=\frac{1}{\sqrt{2c_Lk}}\left(a_L(\vec{k})+a_L^\dagger(-\vec{k})\right)\;,\quad\pi_L(\vec{k})=-i\sqrt{\frac{c_L k}{{2}}}\left(a_L(\vec{k})-a_L^\dagger(-\vec{k})\right)\;,\\
		a_L(\vec{k})&=\frac{1}{\sqrt{2}}\left(\sqrt{c_L k\,}\,\alpha_L(\vec{k})+\frac{i}{\sqrt{c_L k}}\pi_L(\vec{k})\right)\;,\quad a_L^\dagger(\vec{k})=\frac{1}{\sqrt{2}}\left(\sqrt{c_L k\,}\,\alpha_L^\dagger(\vec{k})-\frac{i}{\sqrt{c_L k}}\pi_L^\dagger(\vec{k})\right)\;.
	\end{align}
	In this way, we have the following commutation relations:
	\begin{align}
		[a_T(\vec{k}),a_T^\dagger(\vec{q})]=(2\pi)^3\delta^3(\vec{k}-\vec{q})\;,\quad[a_L(\vec{k}),a_L^\dagger(\vec{q})]=(2\pi)^3\delta^3(\vec{k}-\vec{q})\;.
	\end{align}
	With these conventions, the Hamiltonian becomes:
	\begin{align}
		H=\int \frac{d^3k}{(2\pi)^3}\Bigg(k\,\left(a_T^\dagger(\vec{k})a_T(\vec{k})+\frac{1}{2}[a_T(\vec{k}),a_T^\dagger(\vec{k})]\right)+c_L k\left(a_L^\dagger(\vec{k})a_L(\vec{k})+\frac{1}{2}[a_L(\vec{k}),a_L^\dagger(\vec{k})]\right)\Bigg)\;.
	\end{align}
	While this choice of creation/annihilation operators is convenient for computations, we can note that it makes somewhat cumbersome to describe all the states in the gauge theory limit of $c_L\to 0$. In fact, states with a finite energy due to the presence of a shadow charge (a constant longitudinal electric field without matter) can only be recovered in this limit when the expectation value of the longitudinal number operator $a_L^\dagger a_L$ diverges. This is obtained through the coherent states discussed in Sec.~\ref{sec:shadow}. A different choice in normalization of the operators could have been obtained by choosing different $c_L$ coefficients in Eq.~\eqref{eq:norm}, leading to, \textit{e.g.}, commutators between $a_L$ and $a_L^\dagger$ being proportional to $c_L$, and no appearance of $c_L$ in the Hamiltonian. Such differences in normalization do not change the Hamiltonian as an operator in terms of $\vec{A}$ and $\vec{E}$, and observables remain insensitive with respect to this choice.
	
	We can now define operators in the interaction picture by using the commutators with the Hamiltonian. For example we have:
	\begin{align}
		[a_L(\vec{p}),H]=c_Lp\,a_L(\vec{p})\;,
	\end{align}
	and therefore,
	\begin{align}
		a_L(\vec{p},t)=e^{iHt}a_L(\vec{p})e^{-iHt}=a_L(\vec{p})e^{-ic_Lp\,t}\;.
	\end{align}
	Analogous results follow for the other operators.
	With these choices, the field operators can be expressed as:
	\begin{align}\label{eq:fieldops}
		\!\!\!\!\vec{A}_L(\vec{x},t)=\int \frac{d^3k}{(2\pi)^3}\frac{1}{\sqrt{2c_L k}}\Big(i\hat{k} a_L^\dagger(\vec{k})e^{i(c_L kt-\vec{k}\cdot\vec{x})}-i\hat{k}a_L(\vec{k})e^{-i(c_L kt-\vec{k}\cdot\vec{x})}\Big)\;,\quad\vec{E}_L(\vec{x},t)=\frac{d}{dt}\vec{A}_L(\vec{x},t)\;,\\
		\!\!\!\!\!\!\!\!\vec{A}_T(\vec{x},t)=\int \frac{d^3k}{(2\pi)^3}\frac{1}{\sqrt{2 k}}\sum_{T=1,2}\Big(\vec{e}_T(\vec{k}) a_T^\dagger(\vec{k})e^{i( kt-\vec{k}\cdot\vec{x})}+\vec{e}^{\;*}_T(\vec{k})a_T(\vec{k})e^{-i( kt-\vec{k}\cdot\vec{x})}\Big)\;,\quad\vec{E}_T(\vec{x},t)=\frac{d}{dt}\vec{A}_T(\vec{x},t)\;.
	\end{align}
	With these choices, we can compute the Feynman propagator of the longitudinal modes. We find:
	\begin{align}
		\bra{0}\vec{A}_i(\vec{x}_1,t_1)\vec{A}_j(\vec{x}_2,t_2)\ket{0}=\int\frac{d^3pd^3q}{(2\pi)^6}\Big(&\vec{e}_{T,i}(\vec{p})\vec{e}_{T',j}(\vec{q})\frac{1}{2\sqrt{pq}}\bra{0}a_T(\vec{p})a_{T'}^\dagger(-\vec{q})\ket{0}e^{-i(pt_1-qt_2)}\Big.\\\Big.&+\frac{i^2\hat{p}_i\hat{q}_j}{2 c_L\sqrt{pq}}\bra{0}a_L(\vec{p})a_L^\dagger(-\vec{q})\ket{0}e^{-ic_L(pt_1-qt_2)}\Big)e^{i(\vec{p}\cdot\vec{x}_1+\vec{q}\cdot\vec{x}_2)}\nonumber\\\!\!\!
		=\int\frac{d^3p}{(2\pi)^3}\Big(&\frac{\delta_{ij}-\hat{p}_i\hat{p}_j}{2p}e^{-ip(t_1-t_2)+i\vec{p}\cdot(\vec{x}_1-\vec{x}_2)}+\frac{\hat{p}_i\hat{p}_j}{2c_Lp}e^{-ic_Lp(t_1-t_2)+i\vec{p}\cdot(\vec{x}_1-\vec{x}_2)}\Big)\nonumber\;.
	\end{align}
	Therefore, we obtain the Feynman propagator:
	\begin{align}
		\bra{0}T\{\vec{A}_i(\vec{x}_1,t_1)\vec{A}_j(\vec{x}_2,t_2)\}\ket{0}=\int\frac{d^4p}{(2\pi)^4}\Bigg(&\frac{i(\delta_{ij}-\hat{p}_i\hat{p}_j)}{p_0^2-\vec{p}^2+i\epsilon}e^{-ip_0(t_1-t_2)+i\vec{p}\cdot(\vec{x}_1-\vec{x}_2)}\Bigg.\\\Bigg.&+\frac{i\,\hat{p}_i\hat{p}_j}{p_0^2-c_L^2\vec{p}^2+i\epsilon}e^{-ip_0(t_1-t_2)+i\vec{p}\cdot(\vec{x}_1-\vec{x}_2)}\Bigg)\;.\nonumber
	\end{align}
	To derive more explicit rules for the longitudinal modes, it is useful to express $\vec{A}_L$ in terms of the gradient of a scalar field:
	\begin{align}
		\vec{A}_L=\vec{\nabla}\phi\;,\quad\phi(\vec{x},t)=\int \frac{d^3k}{(2\pi)^3}\frac{1}{\sqrt{2c_L k}}\Big( {\varphi^\dagger(\vec{k})}e^{i(c_L kt-\vec{k}\cdot\vec{x})}+{\varphi(\vec{k})}e^{-i(c_L kt-\vec{k}\cdot\vec{x})}\Big)\;,\quad \varphi(\vec{k})=-\frac{a_L(\vec{k})}{k}\;.
	\end{align}
	From here, we see that $\dot{\phi}$ takes the form:
	\begin{align}
		\dot{\phi}(\vec{x},t)= -c_L\int \frac{d^3k}{(2\pi)^3}\frac{1}{\sqrt{2c_L k}}\Big(i {a_L^\dagger(\vec{k})}e^{i(c_L kt-\vec{k}\cdot\vec{x})}-i{a_L(\vec{k})}e^{-i(c_L kt-\vec{k}\cdot\vec{x})}\Big)\;,\quad \dot{\phi}(\vec{k},t)=-c_L\hat{k}\!\cdot\!\vec{A}_L(\vec{k},t)\;.
	\end{align}
	Therefore, we find the following Feynman propagator:
	\begin{align}
		\bra{0}T\{\dot{\phi}(\vec{x}_1,t_1)\dot{\phi}(\vec{x}_2,t_2)\}\ket{0}=\int\frac{d^4k}{(2\pi)^4}\frac{i\, c_L^2}{k_0^2-c_L^2\vec{k}^2+i\epsilon}e^{-ik_0(t_1-t_2)+i\vec{k}\cdot(\vec{x}_1-\vec{x}_2)}\;.
	\end{align}
	
	To complete our definition of Feynman rules, we have to specify a normalization for single particle states of longitudinal excitations. We follow the common normalization choices:
	\begin{align}
		\ket{k_L}={\sqrt{2c_Lk}}a_L^\dagger(\vec{k},t)\ket{0}\;,
	\end{align}
	where $\ket{0}$ is the vacuum state {in the interaction picture}, normalized to 1.
	With this choice, we have:
	\begin{align}
		\bra{p_L}k_L\rangle={2c_L\sqrt{p\,k}}\bra{0}[a_L(\vec{p}),a_L^\dagger(\vec{k})]\ket{0}=2c_L p(2\pi)^3\delta^3(\vec{p}-\vec{k})\;.
	\end{align}
	Correspondingly, in differential transition rates and cross sections, we will use a phase space measure of the form:
	\begin{align}
		\frac{d^3p}{(2\pi)^3}\frac{1}{2c_L p}\;.
	\end{align}
	These choices lead to usual Feynman rules for the longitudinal modes, with the propagator described above.
	From this, we see that cross sections and rates will feature the usual energy factors also for the Lorentz-breaking longitudinal photon modes in the asymptotic states. Such energy factors appear to lead to inverse powers of $c_L$ in the observables of the theory. Moreover, the form of the propagator makes it clear that powers of $\frac{1}{c_L}$ will generically appear in loop diagrams, once the energy integral is performed through the residue theorem.
	
	\section{Ward identities}\label{app:ward}
	In this appendix we discuss how current conservation can be used in the quantized theory to show explicitly that the slow modes have interactions suppressed by their speed $c_L$. In particular, we show that Ward identities can be used to trade polarization vectors of the longitudinal modes, $\hat{k}$, with factors $c_L \delta^\mu_0$, meaning that the longitudinal modes couple as described in Sec.~\ref{sec:decoupling} and Eq.~\eqref{eq:decouple}.
	
	In fact, in gauge theory, we know that shifting the polarization vector of an external photon by an amount proportional to its momentum does not change the scattering amplitude, as a result of the usual Ward identity on the stripped amplitude:
	\begin{align}
		k^\mu\mathcal{M}_\mu=0\;.
	\end{align}
	Similarly, internal photon propagators carrying momentum $k$ can be shifted by an amount proportional to $k_\mu k_\nu$ without changing the scattering amplitude, which corresponds to a gauge fixing in usual gauge theory.
	In our theory, this continues to be the case thanks to two facts. First, the new longitudinal modes couple through the same interaction vertices as the usual photons, and second, the global $U(1)$ symmetry in the matter sector grants the validity of Ward identities, as a result of Schwinger-Dyson equations. 
	
	For instance, consider an amplitude involving $n$ external longitudinal photons with momenta $\vec{k}_{1}\,,\,\dots\,,\,\vec{k}_n$, as well as other external modes. This can be written in terms of longitudinal polarization vectors $\hat{k}_{1}\,,\,\dots\,,\,\hat{k}_n$ contracted with a stripped amplitude $\mathcal{M}_{i_1\dots i_n}$. Since the longitudinal photon couples to matter through the same operator as the transverse photons, the Ward identities due to the global $U(1)$ symmetry imply that the contraction of this stripped amplitude with the four momentum of a given longitudinal mode vanishes, whenever no Schwinger terms arise. Since the four-momentum of the external longitudinal photons are of the form $k^\mu=(c_Lk,\vec{k})$, we obtain:
	\begin{align}
		\frac{k_1^{i_1}}{{k_1}}\mathcal{M}_{i_1\dots i_n}=c_L \mathcal{M}_{0i_2\dots i_n}\;,
	\end{align}
	implying that the amplitude will be suppressed by a power of $c_L$ for each external longitudinal mode. The corresponding transition rate, which will feature a phase space integral enhanced by $\frac{1}{c_L}$ for each external longitudinal photon, will then be suppressed by $c_L^n$, as discussed in Sec.~\ref{sec:weak}.
	
	Similarly, it will be possible to trade the numerator in the propagator of internal longitudinal modes with a numerator proportional to the energy of the mode, which grants regularity when the internal modes have energy comparable with $c_L$ times their momentum or smaller. In fact, handling properly the time ordering, we find that the propagator can be rewritten as:
	\begin{align}
		\frac{i\delta^i_\mu\delta^j_\nu\,\hat{k}_i\hat{k}_j}{k_0^2-c_L^2\vec{k}^2+i\epsilon}\to\frac{ic_L^2\delta^0_\mu\delta^0_\nu}{k_0^2-c_L^2\vec{k}^2+i\epsilon}\;.
	\end{align}
	The reason for the appearance of $c_L^2$ rather than $k_0^2/\vec{k}^2$, as explained below Eq.~\eqref{eq:propag}, is the particular form of the Schwinger terms that arise, which are proportional to $\frac{1}{\vec{k}^2}$.
	Since the handling of Schwinger terms in general amplitudes is not simple from the point of view of Ward identities, we prefer analyzing the theory through the dressing of matter fields discussed in Sec.~\ref{sec:dressing}.
	
	\bibliographystyle{jhep}
	\bibliography{biblio}

\end{document}